\documentclass[preprint,12pt]{aastex}

\begin{document}
 
\newcommand{\TEEXACT}{T_e}
\newcommand{\TCREXACT}{T_{\times}}
\newcommand{\R}{r}
\newcommand{\PR}{p}
\newcommand{\SE}{S_e}
\newcommand{\AREA}{A}
\newcommand{\PI}{\Pi}
\newcommand{\THETAP}{\tilde{\theta}}
\newcommand{\TCR}{T}
\newcommand{\DEPTH}{\delta}
\newcommand{\FO}{f_0}
\newcommand{\TE}{\tau}
\newcommand{\TF}{T_f}
\newcommand{\TOUT}{T_{{\rm out}}}
\newcommand{\TTRAN}{T_{{\rm transit}}}
\newcommand{\TTOTAL}{T_{{\rm tot}}}
\newcommand{\RHO}{\theta}
\newcommand{\ETA}{\eta}
\newcommand{\TCENTER}{t_c}
\newcommand{\ODD}{\Pi}
\newcommand{\BETA}{\beta}
\newcommand{\TAU}{\tau_0}
\newcommand{\FE}{F^e}
\newcommand{\LE}{\lambda^e}
\newcommand{\FL}{F^l}

\bibliographystyle{apj}

\title{Analytic Approximations for Transit Light Curve Observables,
  Uncertainties, and Covariances}

\author{
Joshua A.~Carter\altaffilmark{1},
Jennifer C.~Yee\altaffilmark{2},
Jason Eastman\altaffilmark{2},\\
B.~Scott Gaudi\altaffilmark{2},
Joshua N.~Winn\altaffilmark{1}
}

\altaffiltext{1}{Department of Physics, and Kavli Institute for
  Astrophysics and Space Research, Massachusetts Institute of
  Technology, Cambridge, MA 02139}

\altaffiltext{2}{Department of Astronomy, Ohio State University, 140
  W.~18th Ave., Columbus, OH 43210}

\begin{abstract}

  The light curve of an exoplanetary transit can be used to estimate
  the planetary radius and other parameters of interest. Because
  accurate parameter estimation is a non-analytic and computationally
  intensive problem, it is often useful to have analytic
  approximations for the parameters as well as their uncertainties and
  covariances.  Here we give such formulas, for the case of an
  exoplanet transiting a star with a uniform brightness distribution.
  We also assess the advantages of some relatively uncorrelated
  parameter sets for fitting actual data.  When limb darkening is
  significant, our parameter sets are still useful, although our
  analytic formulas underpredict the covariances and uncertainties.

\end{abstract}

\keywords{methods: analytical --- binaries: eclipsing --- planets and satellites: general}

\section{Introduction}

The transit of an exoplanet across the face of its parent star is an
opportunity to learn a great deal about the planetary system.
Photometric and spectroscopic observations reveal details about the
planetary radius, mass, atmosphere, and orbit, as reviewed recently by
Charbonneau et al.~(2007). Transit light curves, in particular, bear
information about the planetary and stellar radii, the orbital
inclination, and the mean density of the star (Mandel \& Agol 2000,
\citet{seager03}, Gim\'enez 2007). Additional planets in
the system may be detected through gradual changes in the orbital
parameters of the transiting planet (Miralda-Escud\'e 2002,
Heyl \& Gladman 2007), or from a pattern of anomalies in a collection of
midtransit times (Holman \& Murray 2005, Agol et al.~2005, Ford \& Holman 2007).

In general, the parameters of a transiting system and their
uncertainties must be estimated from the photometric data using
numerical methods. For example, many investigators have used
$\chi^2$-minimization schemes such as AMOEBA or the
Levenberg-Marquardt method, along with confidence levels determined by
examining the appropriate surface of constant $\Delta\chi^2$ (see,
e.g., Brown et al.~2001, Alonso et al.~2004) or by bootstrap methods
(e.g., Sato et al.~2005, Winn et al.~2005). More recently it has
become common to use Markov Chain Monte Carlo methods (e.g., Holman et
al.~2006, Winn et al.~2007, Burke et al.~2007). However, even when
numerical algorithms are required for precise answers, it is often
useful to have analytic approximations for the parameters as well as
their uncertainties and covariances.

Analytic approximations can be useful for planning observations. For
example, one may obtain quick answers to questions such as, for which
systems can I expect to obtain the most precise measurement of the
orbital inclination? Or, how many transit light curves will I need to
gather with a particular telescope before the statistical error in the
planetary radius is smaller than the systematic error? Now that nearly
50 transiting planets are known, we enjoy a situation in which a given
night frequently offers more than one observable transit
event. Analytic calculations can help one decide which target is more
fruitfully observed, and are much simpler and quicker than the
alternative of full numerical simulations. Analytic approximations are
also useful for
understanding the parameter degeneracies inherent in
the model, and for constructing relatively uncorrelated parameter sets
that will speed the convergence of optimization algorithms.
Finally, analytic approximations are useful in
order-of-magnitude estimates of the observability of
subtle transit effects, such as transit timing variations, precession-induced
changes in the transit duration, or the asymmetry in the ingress
and egress durations due to a nonzero orbital eccentricity.

Mandel \& Agol~(2005) and Gim\'enez~(2007) have previously given
analytic formulas for the received flux as a function of the relative
separation of the planet and the star, but their aim was to provide
highly accurate formulas, which are too complex for useful analytic
estimates of uncertainties and covariances. Protopapas et al.~(2007)
provided an analytic and differentiable approximation to the transit
light curve, but they were concerned with speeding up the process of
searching for transits in large databases, rather than parameter
estimation. \citet{seager03} presented an approximate
model of a transit light curve with the desired level of simplicity,
but did not provide analytic estimates of uncertainties and
covariances.

This paper is organized as follows. In \S~\ref{sec:model} we present a
simple analytic model for a transit light curve, using a convenient
and intuitive parameterization similar to that of Seager \&
Mallen-Ornelas~(2003). In \S~\ref{sec:fisher}, we derive analytic
approximations for the uncertainties and covariances of the basic
parameters, and in \S~\ref{sec:accuracy} we verify the accuracy of
those approximations through numerical tests. Our model assumes that
the flux measurements are made continuously throughout the transit,
and that stellar limb-darkening is negligible; in
\S~\ref{subsec:cadence} and \S~\ref{subsec:limb_darkening} we check on
the effects of relaxing these assumptions.  In \S~\ref{sec:prop} we
derive some useful expressions for the uncertainties in some
especially interesting or useful ``derived'' parameters, i.e.,
functions of the basic model parameters. In \S~\ref{sec:smaller_corrs}
we present alternative parameter sets that are better suited to
numerical algorithms for parameter estimation utilizing the analytic
formalism given in \S~\ref{sec:fisher}. We compare the correlations
among parameters for various parameter sets that have been
used in the transit literature. Finally, \S~7 gives a summary of
the key results.

\section{Linear approximation to the transit light curve \label{sec:model}}

Imagine a spherical star of radius $R_\star$ with a uniform brightness
and an unocculted flux $\FO$. When a dark, opaque, spherical planet of
radius $R_p$ is in front of the star, at a center-to-center
sky-projected distance of $z R_\star$, the received stellar flux is
$\FE(\R,z,\FO) = \FO (1-\LE(\R,z))$, where
\begin{eqnarray}
 	\LE(\R,z) = \left\{
		\begin{array}{cc}
			0 & 1+\R < z \\
			\frac{1}{\pi}\left(\R^2 \kappa_0+\kappa_1 -\sqrt{\frac{4 z^2-(1+z^2-\R^2)^2}{4}}\right) & 1-\R < z \leq 1+\R \\
			\R^2 & z \leq 1-\R
		\end{array}
	\right.,
\label{eq:exact}
\end{eqnarray}
with $\kappa_1 = \cos^{-1}[(1-\R^2+z^2)/2z]$ and $\kappa_0 = \cos^{-1}
[(\R^2+z^2-1)/2 \R z]$ (Mandel \& Agol 2002). Geometrically, $\LE$ is
the overlap area between two circles with radii 1 and $r$ whose
centers are $z$ units apart. The approximation of uniform brightness
(no limb darkening) is valid for mid-infrared bandpasses, which are
increasingly being used for transit observations (see, e.g.,
Harrington et al.~2007, Knutson et al.~2007, Deming et al.~2007), and
is a good approximation even for near-infrared and far-red
bandpasses. We make this approximation throughout this paper, except
in \S~\ref{subsec:limb_darkening} where we consider the effect of limb
darkening.

For a planet on a circular orbit, the relation between $z$ and the
time $t$ is
\begin{eqnarray}
z(t) & = & a R_{\star}^{-1}\sqrt{[\sin~n (t-\TCENTER)]^2+[\cos~i \cos~n (t-\TCENTER)]^2}
\label{eq:zexact}
\end{eqnarray}
where $a$ is the semimajor axis, $i$ is the inclination angle, $n
\equiv 2\pi/ P$ is the angular frequency with period $P$, and
$\TCENTER$ is the transit midpoint (when $z$ is smallest).

The four ``contact times'' of the transit are the moments when the
planetary disk and stellar disk are tangent. First contact ($t_{\rm
I}$) occurs at the beginning of the transit, when the disks are
externally tangent. Second contact ($t_{\rm II}$) occurs next, when
the disks are internally tangent. Third and fourth contacts ($t_{\rm
III}$ and $t_{\rm IV}$) are the moments of internal and external
tangency, respectively, as the planetary disk leaves the stellar disk.
The total transit duration is $t_{\rm IV} - t_{\rm I}$. The ingress
phase is defined as the interval between $t_{\rm I}$ and $t_{\rm II}$,
and likewise the egress phase is defined as the interval between
$t_{\rm III}$ and $t_{\rm IV}$. We also find it useful to define the
ingress midpoint $t_{\rm ing} \equiv (t_{\rm I} + t_{\rm II})/2$ and
the egress midpoint $t_{\rm egr} \equiv (t_{\rm III} + t_{\rm IV})/2$.

Although Eqns.~(\ref{eq:exact}) and (\ref{eq:zexact}) give an exact
solution, they are too complicated for an analytic error analysis. We
make a few approximations to enable such an analysis. First, we assume
the orbital period is large compared to transit duration, in which
case Eqn.~(\ref{eq:zexact}) is well-approximated by
\begin{eqnarray}
z(t) & = & \sqrt{[(t-\TCENTER)/\TAU]^2+b^2},
\label{eq:zapp}
\end{eqnarray}
where, for a circular orbit, $\TAU = R_{\star} P/ 2 \pi a = R_{\star}/ n a $ and $b = a \cos
i/R_{\star}$ is the normalized impact parameter. In this limit, the
planet moves uniformly in a straight line across the stellar disk.
Simple expressions may be derived for two
characteristic timescales of the transit:
\begin{eqnarray}
 t_{\rm egr} - t_{\rm ing} & = & \TAU\left(\sqrt{(1+\R)^2-b^2}+\sqrt{(1-\R)^2-b^2}\right) = 2 \TAU \sqrt{1-b^2}+O(\R^2) \label{eq:tcrexact}\\
 t_{\rm II} - t_{\rm I}    & = & \TAU\left(\sqrt{(1+\R)^2-b^2}-\sqrt{(1-\R)^2-b^2}\right) = 2 \TAU \frac{\R}{\sqrt{1-b^2}}+O(\R^3). \label{eq:teexact}
\end{eqnarray}
It is easy to enlarge the discussion to include eccentric orbits,  by 
replacing $a$ by the planet-star distance at midtransit, and
$n$ by the angular frequency at midtransit:
\begin{eqnarray*}
 a      & \rightarrow & \frac{a(1-e^2)}{1 + e\sin\omega}, \\
 n      & \rightarrow & \frac{n(1 + e\sin\omega)^2}{(1-e^2)^\frac{3}{2}},
\end{eqnarray*}
where $e$ is the eccentricity, and $\omega$ is the argument of
pericenter. Here, too, we approximate the planet's actual motion by
uniform motion across the stellar disk, with a velocity
equal to the actual velocity at midtransit.
Methods for computing these quantities at midtransit are discussed by
\citet{murray}, as well as recent transit-specific studies by
Barnes~(2007), Burke~(2008), Ford et al.~(2008), and \citet{gillon07}.  
We redefine the parameters $\TAU$ and $b$ in this expanded scope as
\begin{eqnarray}
	b &\equiv&\frac{a \cos i}{R_{\star}}\left( \frac{1-e^2}{1 + e\sin\omega} \right) \label{eq:b_def}\\
	\TAU & \equiv & \frac{R_{\star}}{a n} \left( \frac{\sqrt{1-e^2}}{1 + e\sin\omega} \label{eq:tau_def}\right).
\end{eqnarray}
We do not restrict our discussion to circular orbits ($e = 0$) unless otherwise stated.

\begin{figure}[htbp] 
   \centering
   \plotone{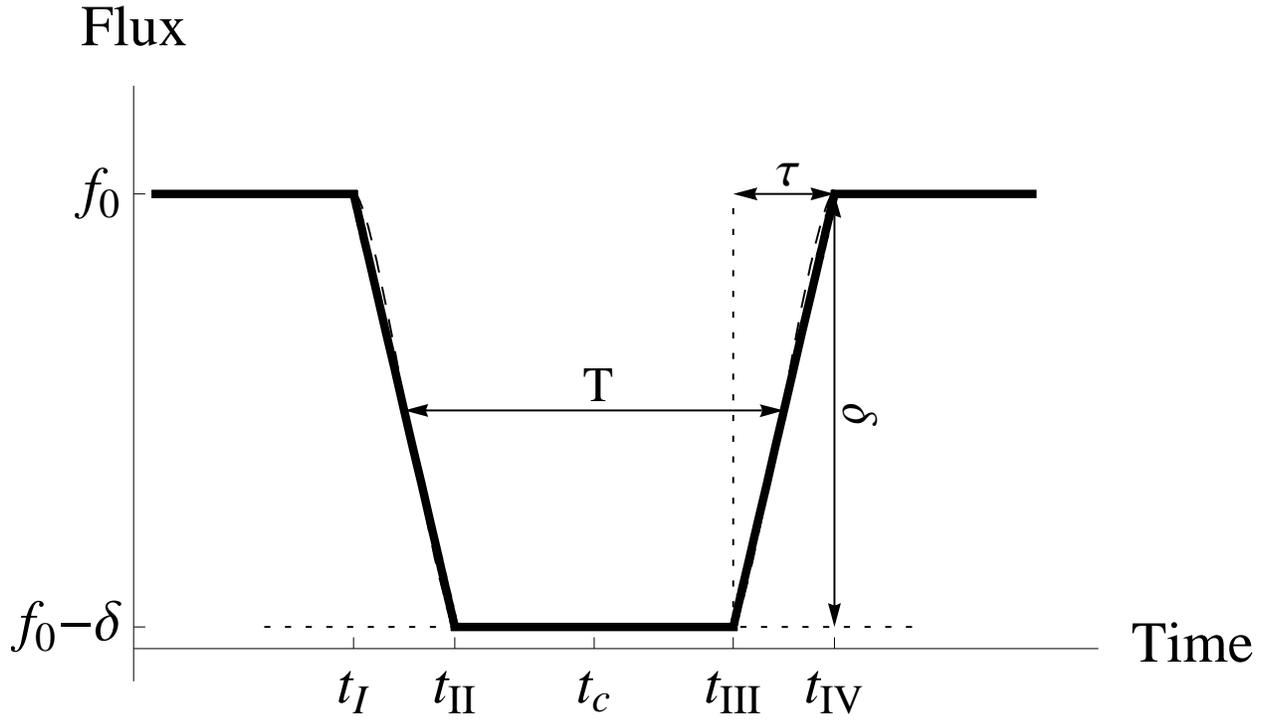}
   \caption{Comparison of the exact and piecewise-linear transit
   models, for the parameter choice $\R = 0.2$, $b = 0.5$. The dashed
   line shows the exact uniform-source model $\FE$, given by Eqn.~(\ref{eq:exact}).
   The solid line shows the linear model $\FL$, given by
   Eqn.~(\ref{eq:linear_model}). }
   \label{fig:model}
\end{figure}

Next, we replace the actual light curve with a model that is
piecewise-linear in time, as illustrated in Figure~\ref{fig:model}.
Specifically, we define the parameters
\begin{eqnarray}
        \DEPTH & \equiv & f_0 r^2 = f_0 (R_p/R_\star)^2 \label{eq:depth} \\
	\TCR   & \equiv & 2 \TAU \sqrt{1-b^2} \label{eq:tcr}\\
	\TE    & \equiv & 2 \TAU \frac{\R}{\sqrt{1-b^2}} \label{eq:te}
\end{eqnarray}
and then we define our model light curve as
\begin{eqnarray}
 	\FL(t) = \left\{
		\begin{array}{cc} 
			\FO - \DEPTH  & |t-\TCENTER| \leq \TCR/2 - \TE/2 \\
			\FO - \DEPTH + \frac{\DEPTH}{\TE} \left( |t-\TCENTER| - \TCR/2 + \TE/2 \right)  & \TCR/2 - \TE/2 < |t-\TCENTER| < \TCR/2 + \TE/2 \\
			\FO & |t-\TCENTER| \geq  \TCR/2 + \TE/2
		\end{array}
	\right.
\label{eq:linear_model}
\end{eqnarray}
We use the symbol $\FL$ to distinguish this piecewise-linear model
($l$ for linear) from the exact uniform-source expression $\FE$
given by Eqns.~(\ref{eq:exact}) and (\ref{eq:zexact}). The deviations
between $\FL$ and $\FE$ occur near and during the ingress and egress
phases. The approximation is most accurate in the limit of small $r$
and $b$ and is least accurate for grazing transits.   As shown in Eqn.~(\ref{eq:teexact}), when $r$ is small, $\TE
\approx t_{\rm II} - t_{\rm I}$ (the ingress or egress duration) and
$\TCR \approx t_{\rm egr} - t_{\rm ing}$ (the total transit
duration). Neither this piecewise-linear model nor the choice of
parameters is new. \citet{seager03} also used a
piecewise-linear model, with different linear combinations of these parameters, and both \citet{burke07} and \citet{bakos07} have
employed parameterizations that are closely related to the parameters
given above. What is specifically new to this paper is an analytic
error and covariance analysis of this linear model, along with useful
analytic expressions for errors in the physical parameters of the
system. The ``inverse'' mapping from our parameterization to a more
physical parameterization is
\begin{eqnarray}
	r^2 = (R_p/R_\star)^2     & = & \DEPTH/\FO \label{eq:mappings_p2}\\
	b^2 = \left(\frac{a \cos i}{R_{\star}}\right)^2\left( \frac{1-e^2}{1 + e\sin\omega} \right)^2 & = & 1-\R \frac{\TCR}{\TE} \label{eq:mappings_b2}\\
 \TAU^2 = \left( \frac{R_{\star}}{a n}\right)^2 \left( \frac{\sqrt{1-e^2}}{1 + e\sin\omega} \right) ^2  & = & \frac{\TCR \TE}{4 \R} \label{eq:mappings_t2}.
\end{eqnarray}

\clearpage
\clearpage

\section{Fisher information analysis \label{sec:fisher}}

Given a model $F(t;\{\PR_i\})$ with independent variable $t$ and a set
of parameters $\{\PR_i\}$, it is possible to estimate the covariance
between parameters, ${\rm Cov}(\PR_i,\PR_j)$, that would be obtained
by measuring $F(t)$ with some specified cadence and
precision. (Gould~2003 gives a pedagogical introduction to this
technique.) Suppose we have $N$ data points taken
at times $t_k$ spanning the entire transit event. The error in each
data point is assumed to be a Gaussian random variable, with zero mean
and standard deviation $\sigma_k$. Then the covariance between
parameters $\{\PR_i\}$ is
\begin{eqnarray}
	{\rm Cov}(p_i,p_j) = \left( B^{-1}\right)_{ij} 
\end{eqnarray}
where $B$ is the zero-mean Gaussian-noise Fisher information matrix, which is calculated as
\begin{eqnarray}
  B_{ij} = \sum_{k=1} ^{N} \sum_{l=1} ^{N}
           \left[ \frac{\partial}{\partial p_i} F(t_k;\{p_m\}) \right]
           {\cal B}_{kl}
           \left[ \frac{\partial}{\partial p_j} F(t_l;\{p_m\}) \right] .
\label{eq:B}
\end{eqnarray} 
Here, ${\cal B}_{kl}$ is the inverse covariance matrix of the flux
measurements. We assume the measurement errors are uncorrelated (i.e.,
we neglect ``red noise''), in which case ${\cal B}_{kl} = \delta_{kl}
\sigma_{k}^{-2}$. We further assume that the measurement errors are
uniform in time with $\sigma_k = \sigma$, giving ${\cal B}_{kl} =
\delta_{kl} \sigma^{-2}$.

In Table~(\ref{tab:derivs}), we compute the needed partial derivatives\footnote{In computing these
derivatives we have ignored the dependence of the piecewise boundaries
in Table.~(\ref{tab:derivs}) on the parameter values.  The
derivatives associated with those boundary changes are finite, and
have a domain of measure zero in the limit of continuous sampling.
Thus they do not affect our covariance calculation.} of the
piecewise-linear light curve $\FL$, which has five parameters $\{p_i\}
= \{\TCENTER, \TE, \TCR, \DEPTH, \FO\}$.
\clearpage
\begin{table}[htbp]
\centering
\begin{tabular}{@{}l|ccc@{}}\hline
	& Totality & Ingress/Egress & Out of Transit \\ \hline
	$\frac{\partial}{\partial \TCENTER} \FL(t;\{\PR_m\})$ & $0$  & $-\frac{\DEPTH}{\TE} \frac{t-\TCENTER}{|t-\TCENTER|}$  & $0$ \\
		$\frac{\partial}{\partial \TE}      \FL(t;\{\PR_m\})$ & $0$  & $-\frac{\DEPTH}{\TE^2}\left( |t-\TCENTER|-\frac{\TCR}{2} \right)$ & $0$ \\
		$\frac{\partial}{\partial \TCR}     \FL(t;\{\PR_m\})$ & $0$  & $-\frac{\DEPTH}{2 \TE}$& $0$ \\
		$\frac{\partial}{\partial \DEPTH}   \FL(t;\{\PR_m\})$ & $-1$ &$ \frac{1}{\TE} \left( |t-\TCENTER|-\frac{\TCR}{2} \right)-\frac{1}{2}$& $0$ \\
		$\frac{\partial}{\partial \FO}      \FL(t;\{\PR_m\})$ & $1$  & $1$ & $1$ \\ \hline
\end{tabular}
\caption{Table of partial derivatives of the piecewise-linear light curve $\FL$, in the five parameters $\{p_i\}
= \{\TCENTER, \TE, \TCR, \DEPTH, \FO\}$.  The intervals $|t-\TCENTER| < \TCR/2-\TE/2$, $\TCR/2-\TE/2 < |t-\TCENTER| < \TCR/2+\TE/2$, and $|t-\TCENTER| > \TCR/2+\TE/2$ correspond to totality, ingress/egress, and out of transit respectively.}\label{tab:derivs}
\end{table}
\clearpage

\begin{figure}[htbp] 
    \centering
     \plotone{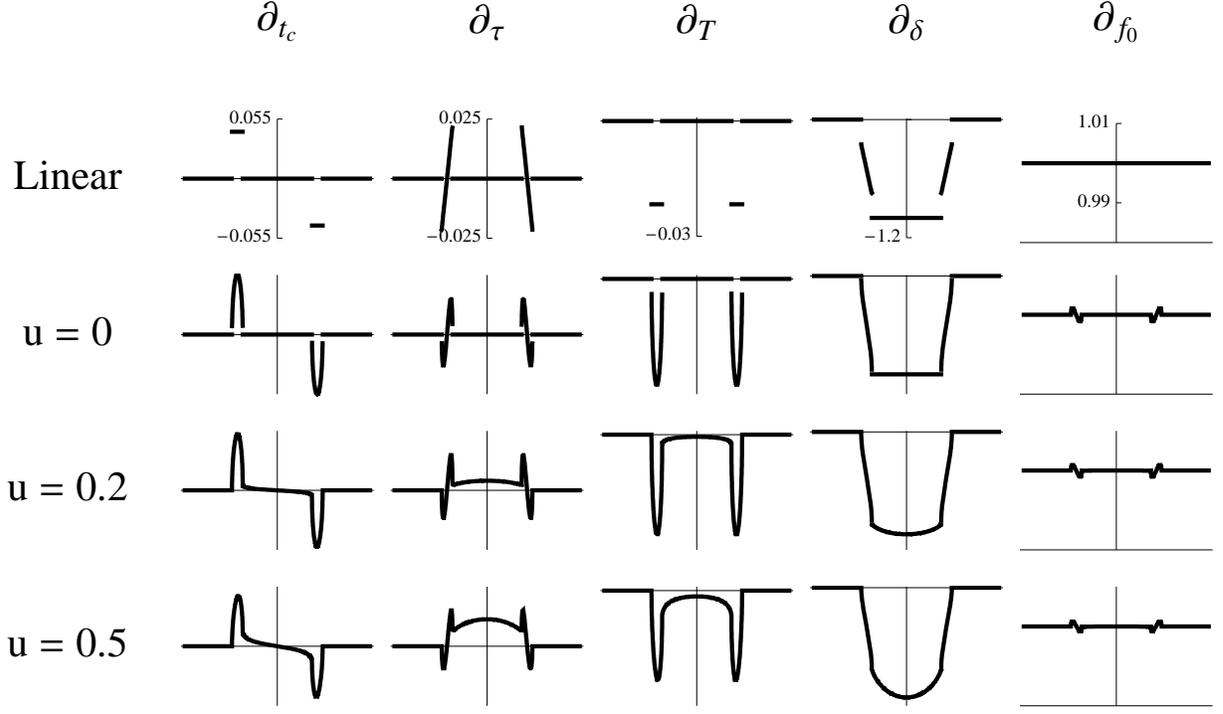}
    \caption{Parameter derivatives, as a function of time, for the
    piecewise-linear model light curve $\FL$ (top row), the exact
    light curve for the case of zero limb darkening $\FE$ (second
    row), and for numerical limb-darkened light curves with a linear
    limb-darkening coefficient $u=0.2$ (third row) and
    $u=0.5$ (bottom row). See \S~\ref{subsec:limb_darkening} for the definition of $u$.  Typical scales are shown in the first row and are consistent in the following rows.}
    \label{fig:model_derivs}
 \end{figure}

Fig.~(\ref{fig:model_derivs}) shows the time dependence of the
parameter derivatives, for a particular case. The time dependence of
the parameter derivatives for the exact uniform-source model $\FE$ is
also shown, for comparison, as are the numerical derivatives for
limb-darkenened light curves. This comparison shows that the linear
model captures the essential features of more realistic models, and in
particular the symmetries. The most obvious problem with the linear
model is that it gives a poor description of the $\TE$-derivative and the $\DEPTH$-derivative for
the case of appreciable limb darkening, as discussed further in
\S~\ref{subsec:limb_darkening}. From Fig.~(\ref{fig:model_derivs})
and Table~(\ref{tab:derivs}) we see that for the parameters
$\TCR$, $\TE$, and $\DEPTH$, the derivatives are symmetric about
$t=\TCENTER$, while the derivative for the parameter $\TCENTER$ is
antisymmetric about $\TCENTER$. This implies that $\TCENTER$ is
uncorrelated with the other parameters. (This is also the case for the
exact model, with or without limb darkening.)

\clearpage
\clearpage

We suppose that the data points are sampled uniformly in time at a
rate $\Gamma = N/\TTOTAL$, where the observations range from $t=t_0$
to $t=t_0+\TTOTAL$ and encompass the entire transit event. In the
limit of large $\Gamma \TE$ we may approximate the sums of
Eqn.~(\ref{eq:B}) with time integrals,
\begin{eqnarray}
	B_{ij} &=& 
          \frac{\Gamma}{\sigma^2}
             \int_{t_0}^{t_0+\TTOTAL}
               \left[ \frac{\partial}{\partial \PR_i} \FL(t;\{\PR_m\}) \right]
               \left[ \frac{\partial}{\partial \PR_j} \FL(t;\{\PR_m\}) \right] ~dt .
	\label{eq:Bint}
\end{eqnarray}
Using the derivatives from Table~(\ref{tab:derivs}) we find
\begin{eqnarray}
	B & = & \frac{\Gamma}{\sigma^2} \left(
\begin{array}{lllll}
 \frac{2 \DEPTH ^2}{\TE} & 0 & 0 & 0 & 0 \\
 0 & \frac{\DEPTH ^2}{6 \TE} & 0 & -\frac{\DEPTH }{6} & 0 \\
 0 & 0 & \frac{\DEPTH ^2}{2 \TE} & \frac{\DEPTH }{2} & -\DEPTH  \\
 0 & -\frac{\DEPTH }{6} & \frac{\DEPTH }{2} & \TCR-\frac{\TE}{3} & -\TCR
   \\
 0 & 0 & -\DEPTH  & -\TCR & \TTOTAL
\end{array}
\right).
\end{eqnarray} 

In what follows, it is useful to define some dimensionless
variables:
\begin{eqnarray}
   Q   & \equiv & \sqrt{\Gamma \TCR} \frac{\DEPTH}{\sigma}, \nonumber \\
 \RHO  & \equiv & \TE/\TCR, \nonumber \\
 \ETA  & \equiv & \TCR/(\TTOTAL - \TCR - \TE).
\label{eq:dimensionless_vars}
\end{eqnarray}
The first of these variables, $Q$, is equal to the total signal-to-noise
ratio of the transit in the limit $r\rightarrow 0$.  The second
variable, $\RHO$, is approximately the ratio of ingress (or egress)
duration to the total transit duration.  The third variable, $\ETA$,
is approximately the ratio of the number of data points obtained
during the transit to the number of data points obtained before or
after the transit. Oftentimes, $r$ and $\RHO$ are much smaller than
unity, which will later enable us to derive simple expressions for the
variances and covariances, but for the moment we consider the general
case.

Inverting $B$, we find the covariance matrix for the piecewise-linear
model,
\begin{eqnarray}
{\rm Cov}(\{\TCENTER, \TE, \TCR, \DEPTH, \FO\}~,\{\TCENTER, \TE, \TCR, \DEPTH, \FO\}) = \;\;\;\;\;\;\;\;\;\;\;\;\;\;\;\;\;\;\;\;\;\;\;\;\;\;\;\;\;\;\;\;\;\;\;\;\;\;\;\;\;\;\;\;\;\;\;\;\;\;\;\;\;\;\;\;\;\;\nonumber \\
 \frac{1}{Q^2} \left(
\begin{array}{lllll}
 \frac{\RHO}{2}  \TCR^2 & 0 & 0 & 0 & 0 \\
 0 & \left[\ETA \RHO+ \frac{6-5 \RHO }{1-\RHO } \right]\RHO  \TCR^2&
   \left[\ETA-\frac{1}{1-\RHO }\right] \RHO ^2 \TCR^2& \left[ \ETA +\frac{1}{1-\RHO }\right]  \RHO \DEPTH \TCR &  \ETA  \RHO  \DEPTH  \TCR \\
 0 &  \left[\ETA-\frac{1}{1-\RHO }\right] \RHO ^2 \TCR^2& \left[ \ETA \RHO+ \frac{2-\RHO}{1-\RHO} \right] \RHO \TCR^2& \left[  \ETA  -\frac{1 }{1-\RHO }\right] \RHO \DEPTH  \TCR&   \ETA  \RHO  \DEPTH \TCR  \\
 0 & \left[ \ETA +\frac{1}{1-\RHO }\right]  \RHO \DEPTH \TCR&\left[  \ETA  -\frac{1 }{1-\RHO }\right] \RHO \DEPTH  \TCR& \left[ \ETA + \frac{1}{1-\RHO}\right]\DEPTH ^2
   &  \ETA  \DEPTH ^2 \\
 0 &   \ETA  \RHO  \DEPTH \TCR &   \ETA  \RHO \DEPTH \TCR &  \ETA \DEPTH ^2
    &  \ETA \DEPTH ^2
\end{array}
\right).\label{eq:cov}
\end{eqnarray}
The elements along the diagonal of the covariance matrix are
variances, or squares of standard errors, $\sigma_{\PR_i} = \sqrt{{\rm
Cov}(\PR_i,\PR_i)}$.

This result can be simplified for the case when many out-of-transit
observations are obtained and $\ETA \rightarrow 0$.  In this limit,
$\FO$ is known with negligible error, and we may assume $\FO =1$
without loss of generality. In this case, $\delta$ is the fractional
transit depth, and the covariance matrix becomes
\begin{eqnarray}
{\rm Cov}(\{\TCENTER, \TE, \TCR, \DEPTH\}~,\{\TCENTER, \TE, \TCR, \DEPTH\}) &=& \frac{1}{Q^2} \left(
\begin{array}{llll}
 \frac{\RHO}{2}   \TCR^2 & 0 & 0 & 0\\
 0 & \frac{\RHO  \left(6-5 \RHO \right) }{1-\RHO } \TCR^2&
   -\frac{ \RHO ^2 }{1-\RHO }\TCR^2& \frac{ \RHO  }{1-\RHO }\DEPTH \TCR \\
 0 &   -\frac{ \RHO ^2 }{1-\RHO } \TCR^2& \frac{\RHO  \left(2-\RHO \right)}{1-\RHO}  \TCR^2& -\frac{  \RHO }{1-\RHO }\DEPTH  \TCR \\
 0 & \frac{ \RHO }{1-\RHO } \DEPTH  \TCR& -\frac{   \RHO  }{1-\RHO }\DEPTH \TCR& \frac{ 1}{1-\RHO}\DEPTH ^2 
\end{array}
\right).\label{eq:sub_cov}
\end{eqnarray}
from which it is obvious that $\RHO$ is the key controlling parameter
that deserves special attention. Using Eqns.~(\ref{eq:tcr}) and
(\ref{eq:te}) we may write
\begin{eqnarray}
	\RHO &= & \frac{\R}{1-b^2}.
\label{eq:RHO}
\end{eqnarray} 

\begin{figure}[htbp] 
   \centering
    \plotone{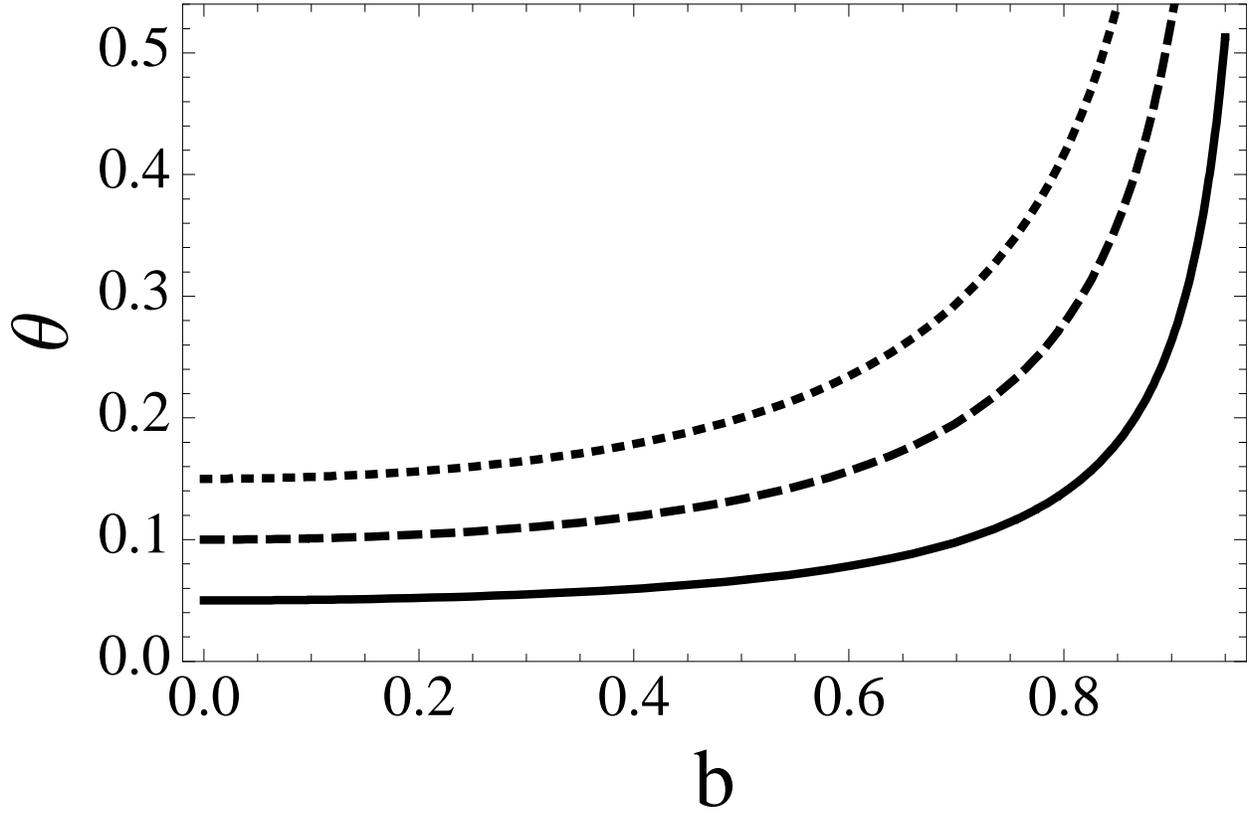}
   \caption{Dependence of $\RHO = \frac{\TE}{\TCR}$ on depth $\DEPTH =
   r^2$ and normalized impact parameter $b$, for the cases $r = 0.05$
   (solid line), $r = 0.1$ (dashed line), and $r=0.15$ (dotted line).}
   \label{fig:rho_in_b}
\end{figure}

\begin{figure}[htbp] 
   \centering
   \plotone{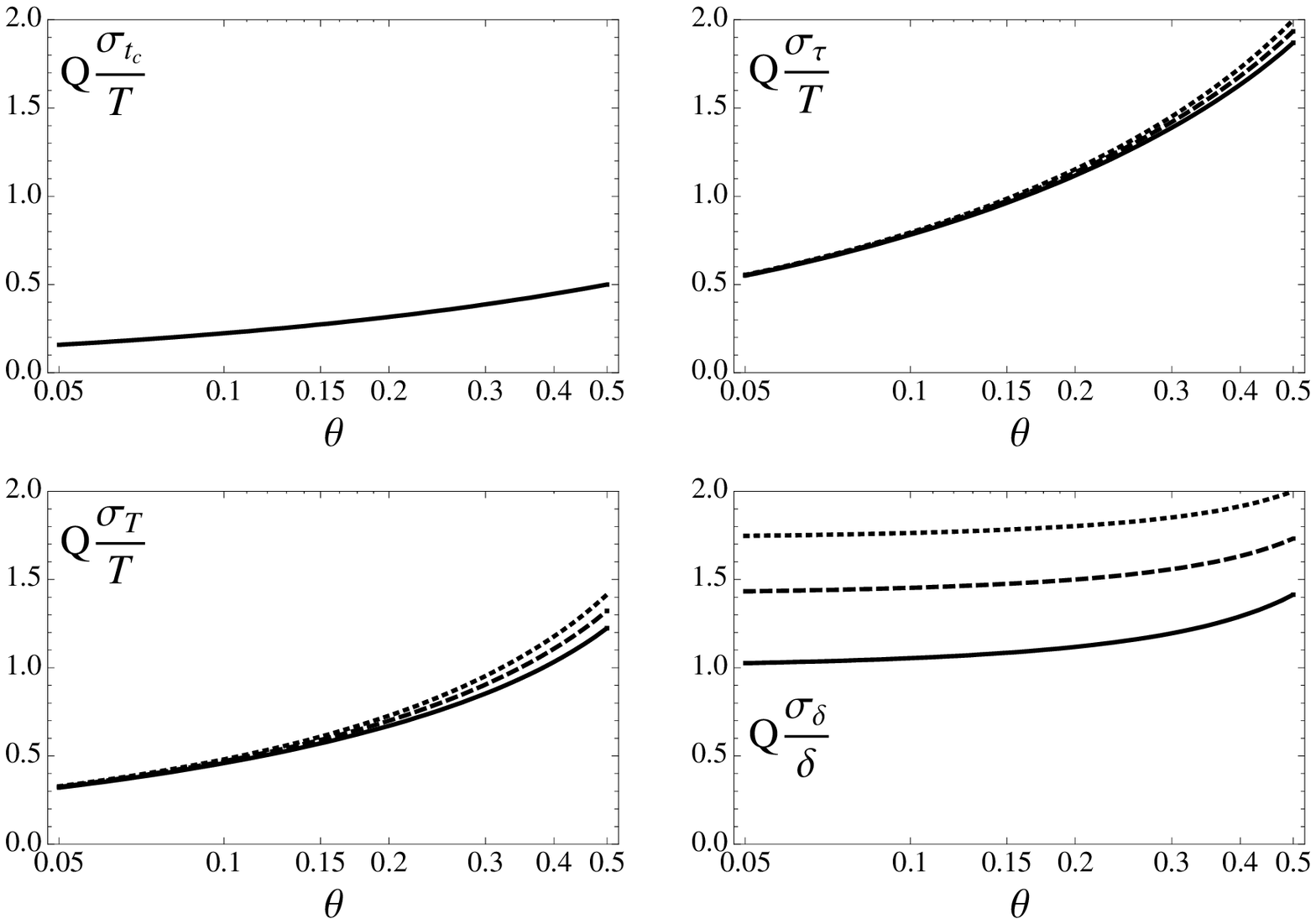}
    \plotone{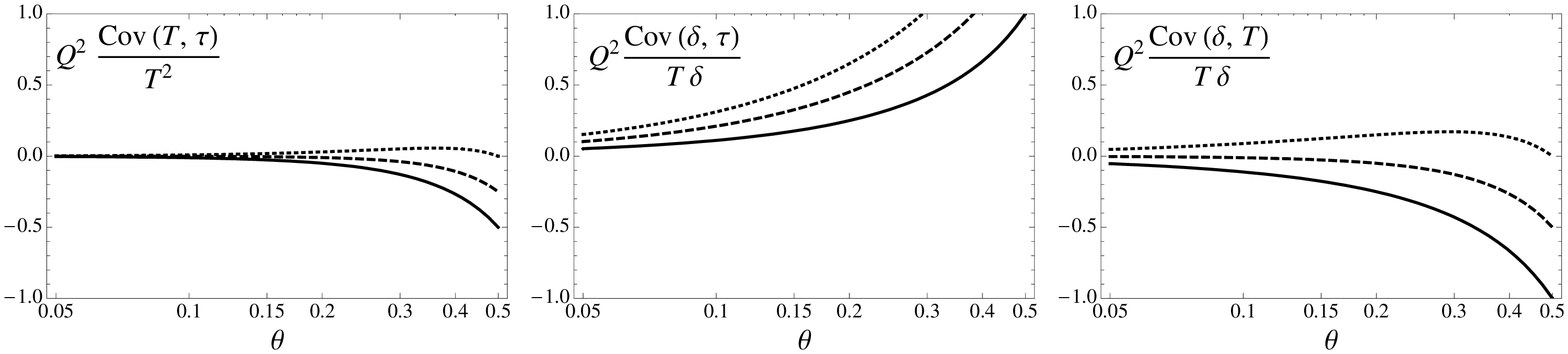}
   \caption{Standard errors and covariances, as a function of $\RHO
   \equiv \TE/\TCR$, for different choices of $\ETA$. The analytic
   expressions are given in Eqn.~(\ref{eq:cov}). The definitions of
   $\ETA$, $\RHO$, and $Q$ are given in
   Eqn.~(\ref{eq:dimensionless_vars}).  Solid line $-$ $\ETA = 0$;
   Dashed line $-$ $\ETA = 0.5$; Dotted line $-$ $\ETA = 1$. }
   \label{fig:var_in_rho}
\end{figure}

Unless the transit is grazing, we have $b \leq 1-r$, and $\RHO$ is
restricted to the range $\left[r\right.$,$\left.\frac{1}{2-r}\right]$.
Fig.~(\ref{fig:rho_in_b}) shows the dependence of $\RHO$ on the impact
parameter, for various choices of the transit depth. It is important
to keep in mind that for $b \lesssim 0.5$, $\RHO$ is nearly equal to
$r$ and depends weakly on $b$. This implies that $\RHO$ is expected to
be quite small for most transiting systems.  For planetary orbits that
are randomly oriented in space, the expected distribution of $b$ is
uniform, and hence we expect $\RHO \lesssim 0.3$ for $90\%$ of a
random sample of transiting planets with $R_p \le R_{\rm Jup}$\footnote{In fact, the fraction of
  discovered systems with $\RHO \lesssim 0.3$ may be even larger than
  90\%, because selection effects make it harder to detect grazing
  transits.}. For this reason, in the following figures we use a
logarithmic scale for $\RHO$, to emphasize the small values.
Fig.~(\ref{fig:var_in_rho}) shows the (suitably normalized) elements
of the covariance matrix as a function of $\RHO$.

\clearpage

\clearpage
	
In the limits $\ETA\rightarrow 0$ (errorless knowledge of $\FO$) and
$\RHO\rightarrow r$ (small impact parameter), the expressions for the
standard errors are especially simple:
\begin{eqnarray}
	\sigma_{\TCENTER} & = & Q^{-1} \TCR \sqrt{\RHO/2},  \nonumber\\
	\sigma_{\TE}      & \approx & Q^{-1} \TCR \sqrt{6 \RHO},  \nonumber\\
	\sigma_{\TCR}     & \approx & Q^{-1} \TCR \sqrt{2 \RHO},  \nonumber \\
	\sigma_{\DEPTH}   & \approx & Q^{-1} \DEPTH .
\label{eq:covlimits}
\end{eqnarray}
In this regime, we have a clear hierarchy in the precision with which
the time parameters are known, with $\sigma_{\TCENTER} <
\sigma_{\TCR} < \sigma_{\TE}$.

To further quantify the degree of correlation among the parameters, we
compute the correlation matrix,
\begin{eqnarray}
{\rm Corr}(\{\TCENTER, \TE, \TCR, \DEPTH, \FO\}~ ,\{\TCENTER, \TE, \TCR, \DEPTH, \FO\}) = \left\{\frac{{\rm Cov}(i, j)}{\sqrt{ {\rm Cov}(i, i) {\rm Cov}(j, j)}}\right\} =\;\;\;\;\;\;\;\;\;\;\;\;\;\;\;\;\;\;\;\;\;\;\;\;\;\;\;\;\;\;\;\;\;\;\;\;\;\;\;\;\;\;\;\;\;\;\;\;\;\;\;\nonumber \\
	\left(
\begin{array}{lllll}
 1 & 0 & 0 & 0 & 0 \\
 0 & 1 & \frac{(\BETA -1) \RHO }{\sqrt{(6-\RHO(5-\BETA)) (2-\RHO (1-\BETA))}} &\sqrt{ \frac{(\BETA +1) \RHO }{
   6-\RHO(5-\BETA)}} & \sqrt{\frac{\BETA \RHO }{6-\RHO(5-\BETA) }} \\
 0 & \frac{(\BETA -1) \RHO }{\sqrt{(6-\RHO(5-\BETA) ) (2-\RHO (1-\BETA))}} & 1 & \frac{(\BETA -1) \sqrt{\RHO }}{\sqrt{(\BETA +1)
   (2-\RHO(1-\BETA ) )}} & \sqrt{\frac{\BETA \RHO}{ 2-\RHO (1-\BETA)}} \\
 0 & \sqrt{ \frac{(\BETA +1) \RHO }{
   6-\RHO(5-\BETA) }} & \frac{(\BETA -1) \sqrt{\RHO }}{\sqrt{(\BETA +1)
   (2-\RHO(1-\BETA ) )}} & 1 & \sqrt{\frac{\BETA }{ \BETA +1}} \\
 0 &   \sqrt{\frac{\BETA \RHO }{6-\RHO(5-\BETA) }}& \sqrt{\frac{\BETA \RHO}{ 2-\RHO (1-\BETA)}}& \sqrt{\frac{\BETA }{ \BETA +1}} & 1
\end{array}
\right). \label{eq:corr}
\end{eqnarray} 
where we have defined $\BETA \equiv \ETA (1-\RHO)$ to simplify the
resulting expression.  For $\RHO \rightarrow 0$, all correlations with
$\FO$ vanish except for the correlation with $\DEPTH$. Due to the fact the correlation between $\DEPTH$ and $\FO$ is $\propto \BETA^{1/2}$, it remains large even for fairly small $\BETA$.  In the limit of
$\ETA \rightarrow 0$ ($\BETA \rightarrow 0$), we remove all
correlations with $\FO$ and have the remaining correlations
depending only on the ratio $\RHO$:
\begin{eqnarray}
\lim_{\ETA \rightarrow 0}{\rm Corr}(\cdot~ ,\cdot) &=&	\left(
\begin{array}{llllll}
 1 & 0 & 0 & 0  & 0\\
 0 & 1 & -\frac{\RHO }{\sqrt{(6 -5 \RHO) (2- \RHO )}} &\sqrt{ \frac{\RHO }{
   6 -5 \RHO}} &0 \\
 0 &  -\frac{\RHO }{\sqrt{(6 -5 \RHO) (2- \RHO )}} & 1 & -\sqrt{\frac{ \RHO }{
   2-\RHO }} & 0 \\
 0 & \sqrt{ \frac{\RHO }{
   6 -5 \RHO}}  & -\sqrt{\frac{ \RHO }{
   2-\RHO }} & 1  &0 \\
   0& 0&0&0&1
\end{array}
\right). \label{eq:sub_corr}
\end{eqnarray} 
Correlations with $\FO$ decline with $\ETA$ as $\sqrt{\ETA}$. 

\begin{figure}[htbp] 
   \centering
    \plotone{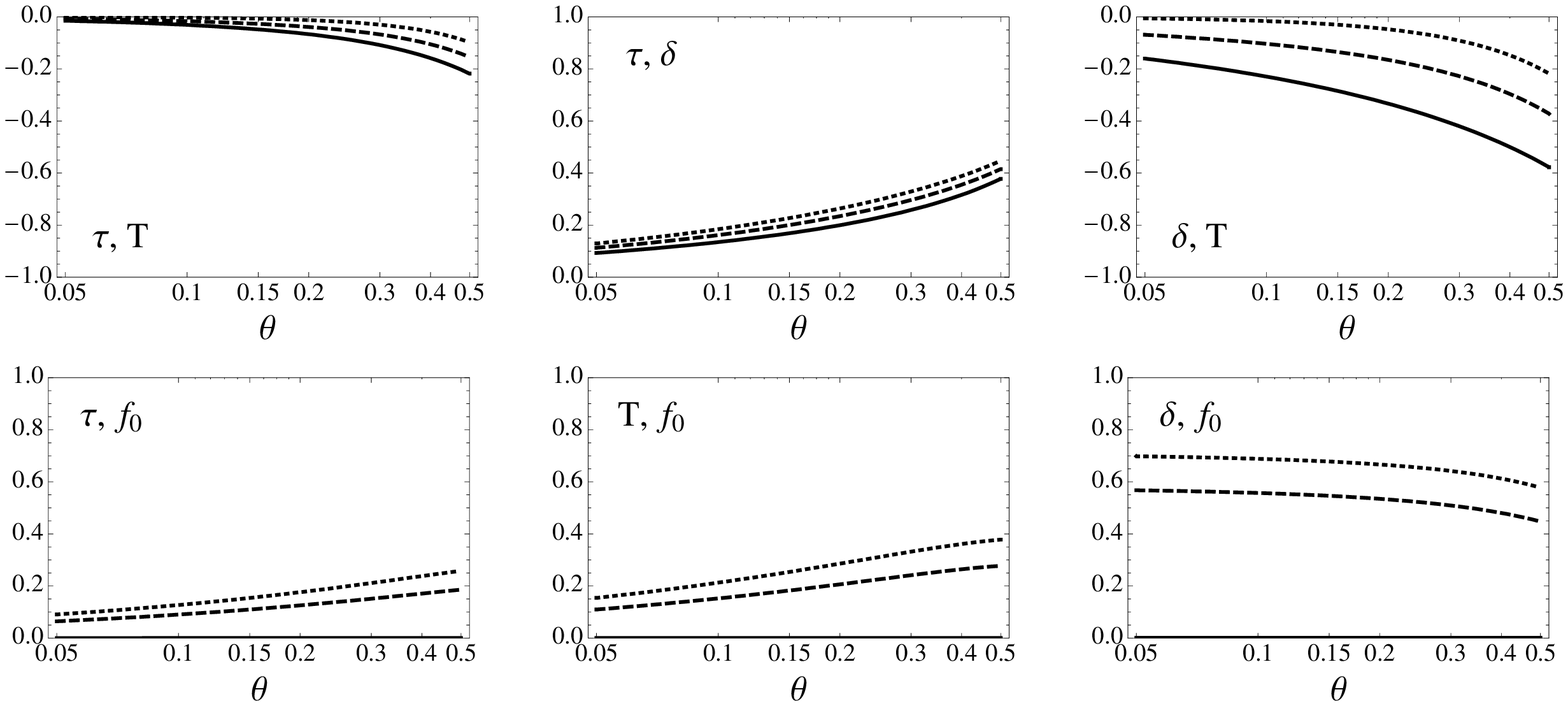}
   \caption{Correlations of the piecewise-linear model parameters, as a function
   of $\RHO \equiv \TE/\TCR$ for different choices of $\ETA$.
   Solid line $-$ $\ETA = 0$; Dashed
   line $-$ $\ETA = 0.5$; Dotted line $-$ $\ETA = 1$.}
   \label{fig:corr_in_rho}
\end{figure}

In Fig. (\ref{fig:corr_in_rho}), we have plotted the nonzero
correlations as a function of $\RHO$ for a few choices of $\ETA$.  The
special case of $\ETA \rightarrow 0$ is plotted in
Fig.~(\ref{fig:corr_in_rho_eta_0}). In the $\ETA \rightarrow 0$ limit,
all correlations are small ($\lesssim 0.3$) over a large region of the
parameter space. Thus, our choice of parameters provides a weakly
correlated set for all but grazing transits ($\RHO \sim 1/2$), as noted
during the numerical analysis of particular systems by Burke et
al.~(2007) and Bakos et al.~(2007). One naturally wonders whether a
different choice of parameters would give even smaller (or even zero)
correlations.  In \S~\ref{sec:smaller_corrs} we present parameter
sets that are essentially uncorrelated and have other desirable properties
for numerical parameter estimation algorithms.

The analytic formalism given in this section and more
specifically the simple analytic covariance matrices in
Eqns.~(\ref{eq:cov}, \ref{eq:sub_cov}) provide a toolbox with which to
evaluate the statistical merits of any parameter set that can be
written in terms of our parameters. In \S~\ref{sec:prop} this
technique is defined and applied to produce analytic formulas for
variances, covariances and uncertainties in several interesting
parameters.  

\clearpage
\newcommand{\corr}[2]{${\rm Corr}(#1,#2)$}
\begin{figure}[htbp] 
   \centering
    \plotone{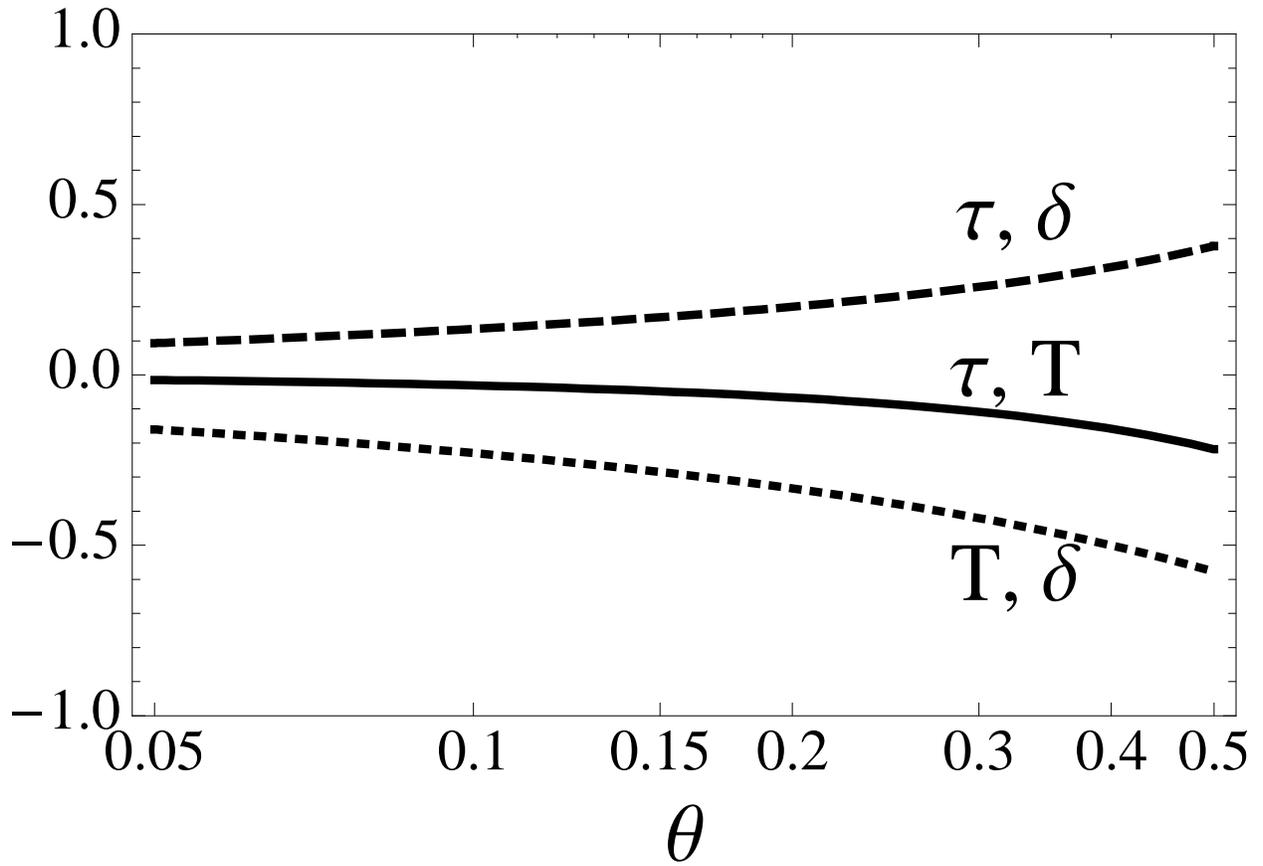}
   \caption{Correlations of the piecewise-linear model parameters,
   as a function of $\RHO \equiv \TE/\TCR$, for the case 
   $\ETA \rightarrow 0$ (errorless knowledge of the out-of-transit flux).
   Solid line $-$ \corr{\TE}{\TCR}. Dashed line $-$ \corr{\TE}{\DEPTH}. Dotted line $-$ \corr{\TCR}{\DEPTH}.}
   \label{fig:corr_in_rho_eta_0}
\end{figure}
\clearpage

\section{Accuracy of the covariance expressions \label{sec:accuracy}}

Before investigating other parameter sets, it is necessary to examine
the validity of Eqns.~(\ref{eq:cov}, \ref{eq:sub_cov}, \ref{eq:corr},
\ref{eq:sub_corr}) when compared to similar quantities derived from
more realistic transit light curve models.
The utility of the covariance matrix in Eqn.~(\ref{eq:cov}) depends on
the accuracy of the integral approximation of Eqn. (\ref{eq:Bint}),
and on the fidelity with which the parameter dependences of the
piecewise-linear model mimic the dependences of the exact uniform-source
model.  In this section we investigate these two issues.

\subsection{Finite cadence correction \label{subsec:cadence}}

The case of a finite observing cadence, rather than continuous
sampling, can be analyzed by evaluating the exact sums of
Eqn.~(\ref{eq:B}). Generally, given a sampling rate $\Gamma$, we
expect the integral approximation in Eqn.~(\ref{eq:Bint}) to be valid
to order $(\Gamma \TE)^{-1}$. In the $\ETA \rightarrow 0$ limit we may
evaluate the exact sums, under the assumption of a uniform sampling
rate, with data points occurring exactly at the start and end of the
ingress (and egress) phases as well as at some intermediate times.
This directly summed covariance, ${\rm Cov}_{\rm sum}$, is related to
the integral-approximation covariance Eqn.~(\ref{eq:sub_cov}) as
\begin{eqnarray}
	{\rm Cov}_{\rm sum}(\cdot,\cdot) &=& {\rm Cov}(\cdot,\cdot) + 6 \left(\frac{\TCR}{Q}\right)^2\frac{ \RHO}{1-\epsilon^2}
	\left(\begin{array}{llll}
		0 & 0 & 0 &0 \\
		0 & \epsilon^2 & \epsilon&0 \\
		0 & \epsilon & \epsilon^2 & 0 \\
		0 & 0 & 0 & 0
	\end{array}\right)  \label{eq:sums}
\end{eqnarray}
where $\epsilon = (\Gamma \TE)^{-1}$.

The quantity $\Gamma \TE$ is approximately the number of data points
obtained during ingress or egress.  It is evident from
Eqn.~(\ref{eq:sums}) that for this sampling scheme only the variances
of $\TCR$ and $\TE$ along with their covariance are corrected.  The
corrections to the variances and covariance are $O(\epsilon^2)$ and
$O(\epsilon)$ respectively.

\subsection{Comparison with covariances of the exact uniform-source model \label{sec:comp}}

We tested the accuracy of the covariance matrix based on the
piecewise-linear model by (1) performing a numerical Fisher analysis
of the exact uniform-source model, and also (2) applying a Markov Chain Monte Carlo
(MCMC) analysis of simulated data based on the exact uniform-source
model.  In both analyses, orbits are assumed to be circular.
  For the first task, we evaluated the analytic parameter
derivatives of Eqn.~(\ref{eq:exact}), which are too cumbersome to be
worth reproducing here, and numerically integrated
Eqn.~(\ref{eq:Bint}) to generate covariance matrices over a wide range
of parameter choices.  Fig.~(\ref{fig:model_derivs}), in \S~2, shows
the parameter derivatives for the exact model, as well as the
piecewise-linear model and some limb-darkened light curves.  For the
second task, idealized data was generated by adding Gaussian noise with
standard deviation $\sigma/\FO = 5\times10^{-4}$ to
Eqn.~(\ref{eq:exact}) sampled at $\Gamma = 100$ (in units of the
characteristic timescale $\TAU$, Eqn.~(\ref{eq:tau_def})). With
this sampling rate, approximately 50 samples occur during the ingress
and egress phases. Approximately $10^4$ links per parameter were generated
with a Gibbs sampler and a Metropolis-Hasting jump-acceptance
criterion. The jump-success fraction (the fraction of jumps in
parameter space that are actually executed) was approximately $25\%$
for all parameters. The effective length, defined as the ratio
of the number of links to the correlation length (see the end of
\S~\ref{sec:smaller_corrs} for the exact definition), was roughly
$1000-2000$ for the piecewise-linear model parameter set. More details
on the MCMC algorithm are given by \citet{tegmark04} and
\citet{ford05}. Standard errors were determined by computing the standard deviation of
the resulting distribution for each parameter. The Fisher-information
analysis should mirror the MCMC results, as long as the log-likelihood
function is well approximated as quadratic near the mean (Gould~2003).

\begin{figure}[htbp] 
   \centering
   \plotone{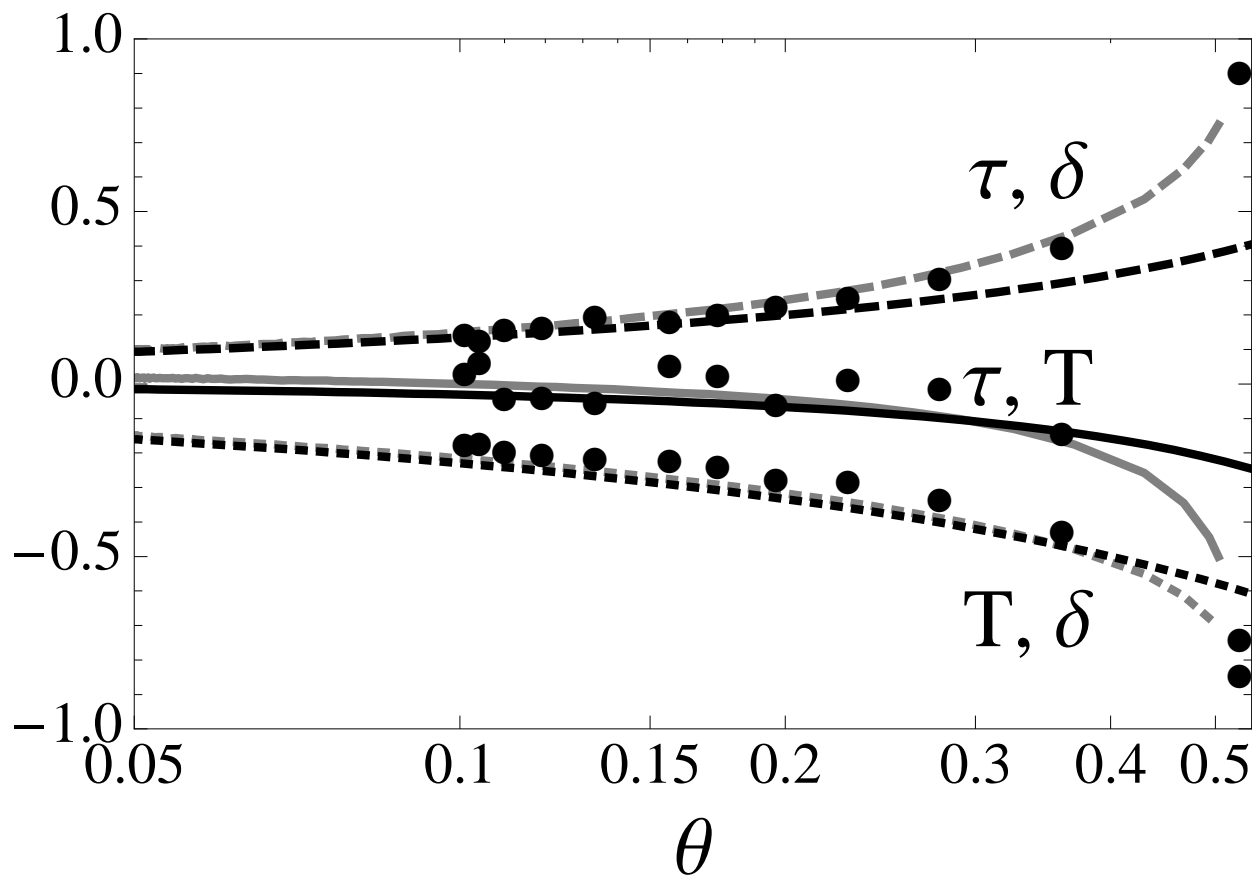} 
   \caption{Comparison of the non-zero correlation matrix elements for
     the exact light-curve model and the piecewise-linear model, as a
     function of $\RHO \equiv \TE/\TCR$, for $\ETA \rightarrow 0$.
     Black curves:~correlations for the piecewise-linear model.  Gray
     curves:~correlations for the exact uniform-source model.  Black
     dots:~correlations based on an MCMC analysis of simulated
     data with Gaussian noise ($r=0.1$).}
   \label{fig:corr_eta_0}
\end{figure}

\begin{figure}[htbp] 
   \centering
    \plotone{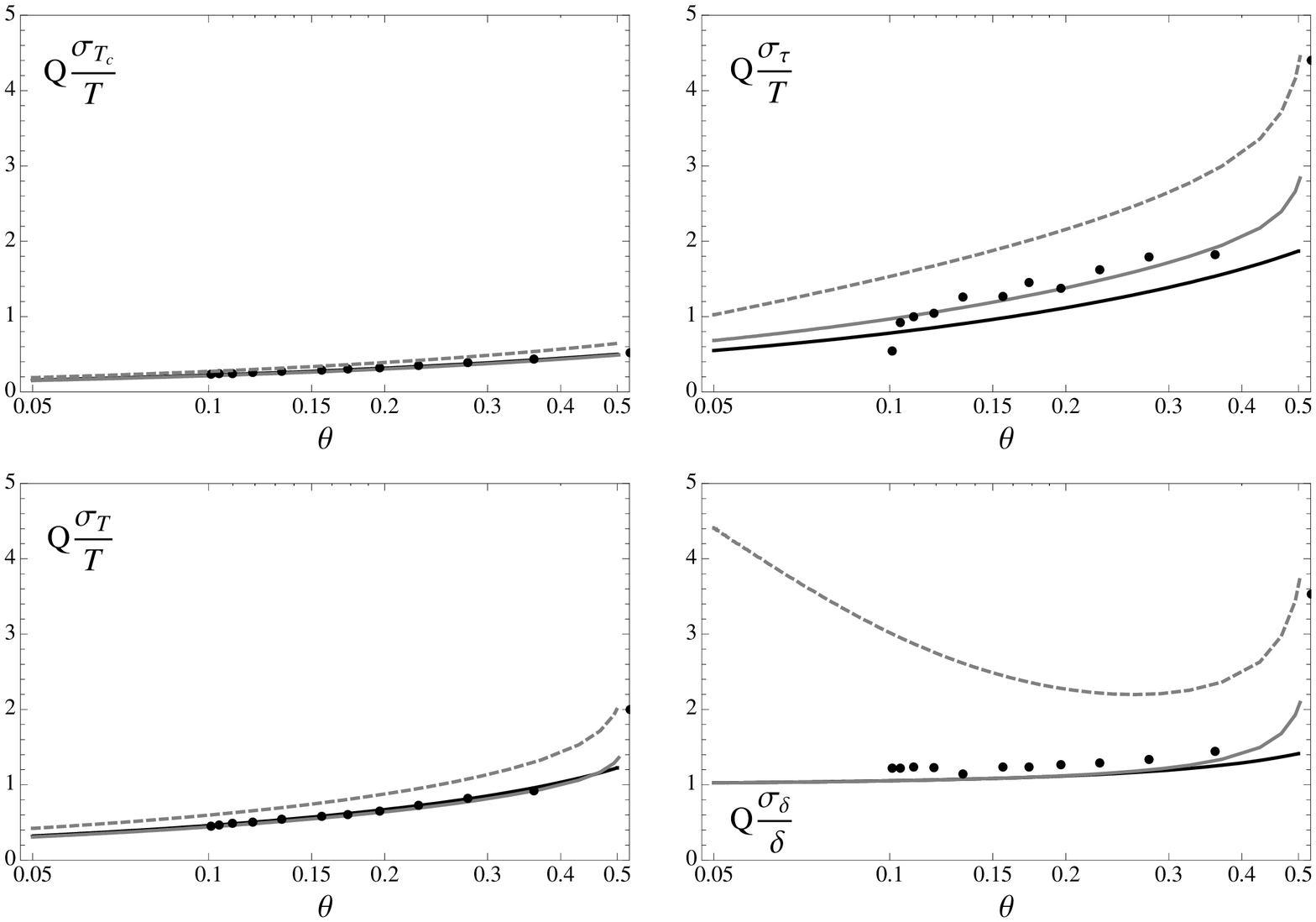}
    \plotone{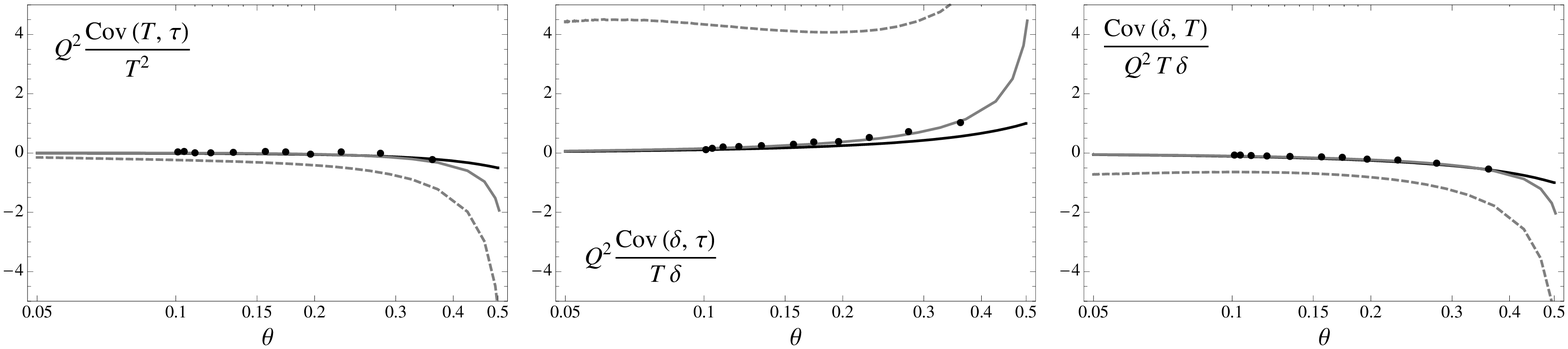}
   \caption{Comparison of the covariance matrix elements for the exact
     uniform-source model, linear limb-darkened model, and the
     piecewise-linear model, as a function of $\RHO \equiv \TE/\TCR$,
     for $\ETA \rightarrow 0$.  Black curves:~covariances for the
     piecewise-linear model.  Gray curves:~covariances for the exact
     model with linear limb-darkening coefficient $u = 0$ (solid) and
     $u = 0.5$ (dashed). Black dots:~covariances as determined by a
     MCMC analysis of simulated data with Gaussian noise ($u = 0$ and
     $r=0.1$).  The dimensionless number $Q \equiv \sqrt{ \Gamma \TCR
     } \DEPTH/\sigma$ (see Eqn.~\ref{eq:dimensionless_vars}) is
     approximately the signal-to-noise ratio of the transit.}
   \label{fig:cov_numerical}
\end{figure}

The numerical Fisher analysis was performed for $\ETA = 0$ and $0.05
\leq \RHO \lesssim 1/2$. In practice this was done by choosing $\R =
0.05$ and varying $b$ across the full range of impact parameters. (The
numerical analysis confirmed that the suitably-normalized covariances
vary only as a function of $\RHO \equiv \TE/\TCR$, with the exception
of slight $\DEPTH$-dependent positive offset in $\sigma_\DEPTH$ that
goes to zero as $\DEPTH$ goes to zero.) The MCMC analysis for $\ETA = 0$ was
accomplished by fixing the out of transit flux, $\FO = 1$, and varying
the remaining parameters We chose $\R = 0.1$ for the MCMC analysis.
Fig.~(\ref{fig:corr_eta_0}) shows all of the nonzero numerical
correlation matrix elements, as a function of $\RHO$.  The MCMC
results, plotted as solid symbols, closely follow the curves resulting
from the numerical Fisher analysis.  Fig.~(\ref{fig:cov_numerical})
shows the nonzero numerical covariance matrix elements, also for the
case $\ETA=0$.

The correlations of the piecewise-linear model match the correlations
of the exact model reasonably well, with the most significant
deviations occurring only in the grazing limit, $\RHO \sim 1/2$.  We
have also confirmed that a similar level of agreement is obtained for
nonzero $\ETA$, although for brevity those results are not shown here.
We concluded from these tests that the errors in the analytic
estimates of the uncertainties are generally small enough for the
analytic error estimates derived from the piecewise-linear model to be
useful.

\clearpage

\clearpage

\subsection{The effects of limb darkening \label{subsec:limb_darkening}}

The piecewise-linear function of Eqn.~(\ref{eq:linear_model}) was
constructed as a model of a transit across a stellar disk of uniform
brightness, with applications to far-red and infrared photometry in
mind. At shorter wavelengths, the limb darkening of the star is
important. How useful are the previously derived results for this
case, if at all? We used the limb-darkened light-curve models given by
\citet{agol02} to answer this question.

To simplify the analysis we adopted a ``linear'' limb-darkening
law, in which the surface brightness profile of the star is
\begin{eqnarray}
	\frac{I(z)}{I_0} & = & 1-u \left( 1- \sqrt{1-z^2} \right)
\end{eqnarray}
where $u$ is the linear limb-darkening parameter. Claret (2000)
finds values of $u$ ranging from $0.5$--$1.2$ in $UBVR$ for a
range of main-sequence stars.  Longer wavelength bands correspond to a
smaller $u$ for the same surface gravity and effective
temperature.  Solar values are $u \approx 0.5$ in the Johnson
$R$ band and 0.2 in the $K$ band. Fig.~(\ref{fig:model_derivs}) of
\S~2 shows the time-dependence of the parameter derivatives of a
linear limb-darkened light curve, for the two cases $u=0.2$ and
$u=0.5$, to allow for comparison with the corresponding
dependences of the piecewise-linear model and the exact model with no
limb darkening. 

\begin{figure}[htbp] 
   \plottwo{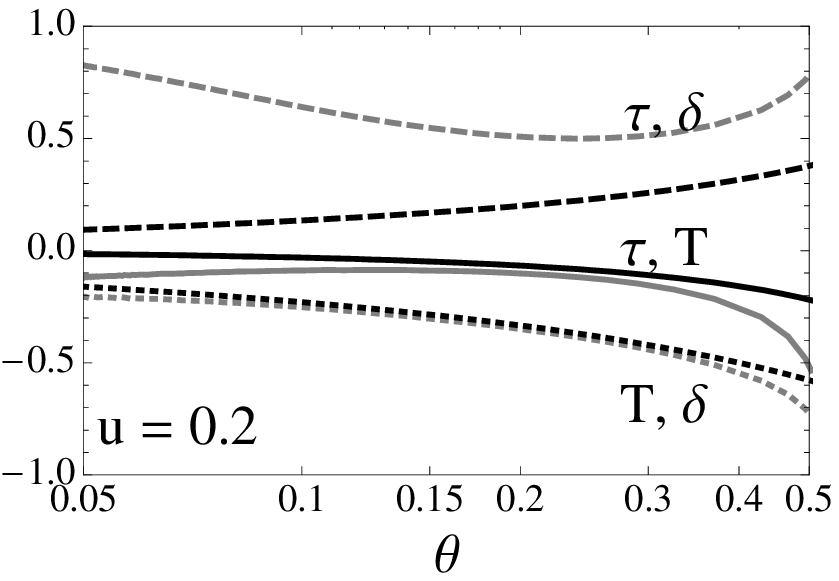}{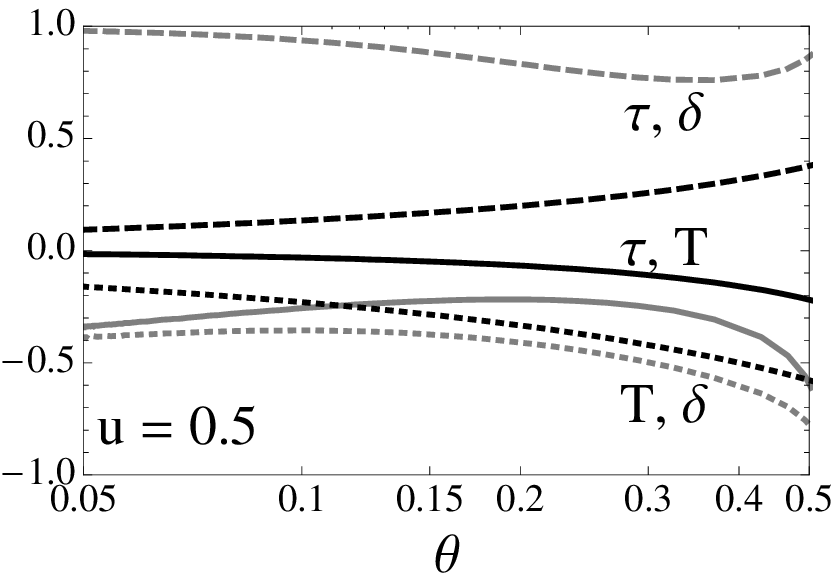}
  \caption{Comparison of the analytic correlations (black lines;
   Eqn.~\ref{eq:corr}) numerically-calculated correlation matrix
   elements for a linear limb-darkened light curve (gray lines), as a
   function of $\RHO \equiv \TE/\TCR$, for $\ETA\rightarrow 0$.  Linestyles follow the conventions of Fig.~(\ref{fig:corr_eta_0}).}
   \label{fig:corr_ld_no_fix}
\end{figure}

From the differences apparent in this plot, one would expect increased
correlations (larger than our analytic formulas would predict) between
the transit depth and the two timescales $\TE$ and $\TCR$. This is
borne out by our numerical calculations of the covariance matrix
elements, which are plotted in
Figs.~(\ref{fig:cov_numerical},\ref{fig:corr_ld_no_fix}).  The
analytic formulas underpredict the variances in $\DEPTH$ and $\TE$ by
a factor of a few, and they also severely underpredict the correlation
between those parameters.  

\clearpage
\clearpage

\begin{figure}[htbp] 
   \centering
    \plotone{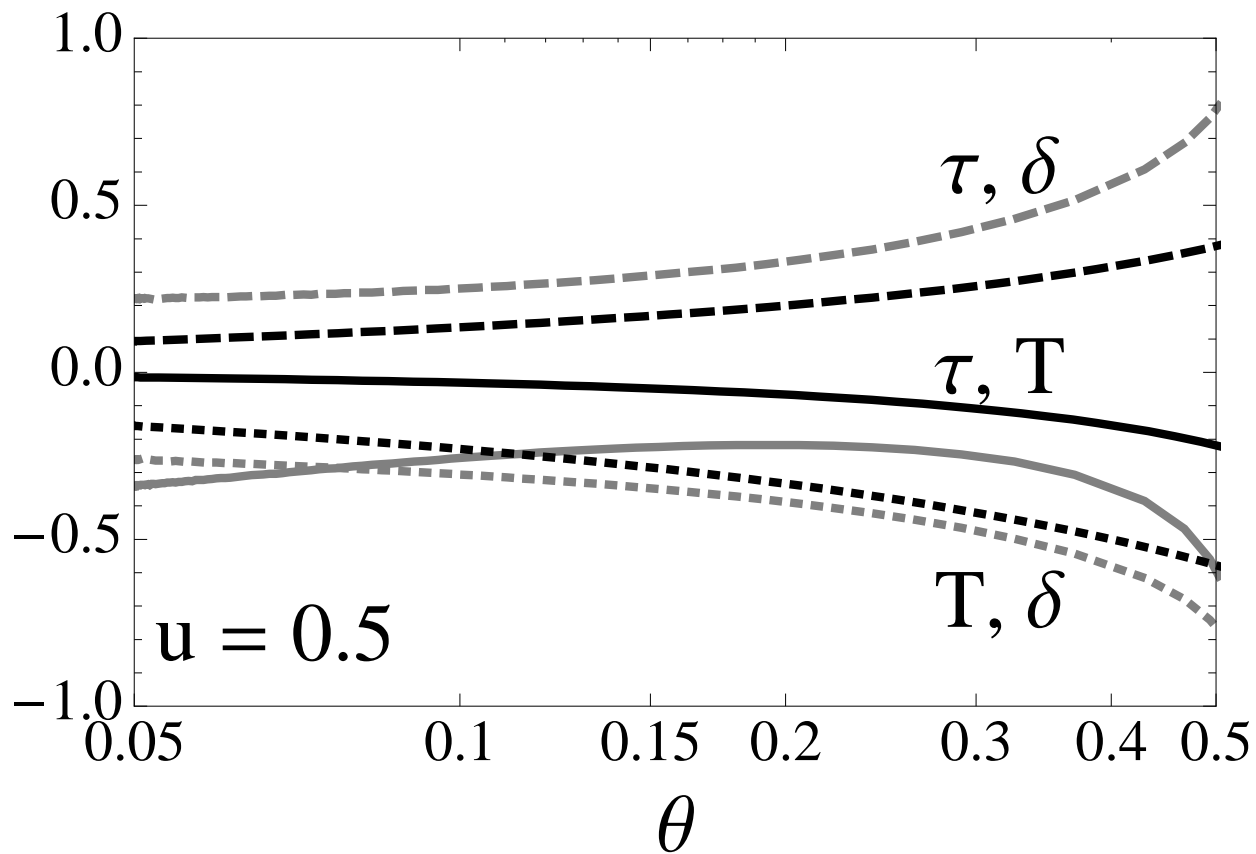}
   \caption{Comparison of correlation matrix elements for the
     piecewise-linear model (black curve) and a linear limb-darkened
     light curve ($u=0.5$; gray curve), as a function of $\RHO
     \equiv \TE/\TCR$.  Here, the $\DEPTH$ parameter has been
     redefined as the minimum of the limb-darkened light curve, as
     approximated by Eqn.~(\ref{eq:depth_ld}). Linestyles follow the conventions of Fig.~(\ref{fig:corr_eta_0}).}
   \label{fig:corr_ld_05_fix} 
\end{figure}

\begin{figure}[htbp] 
   \centering
    \plotone{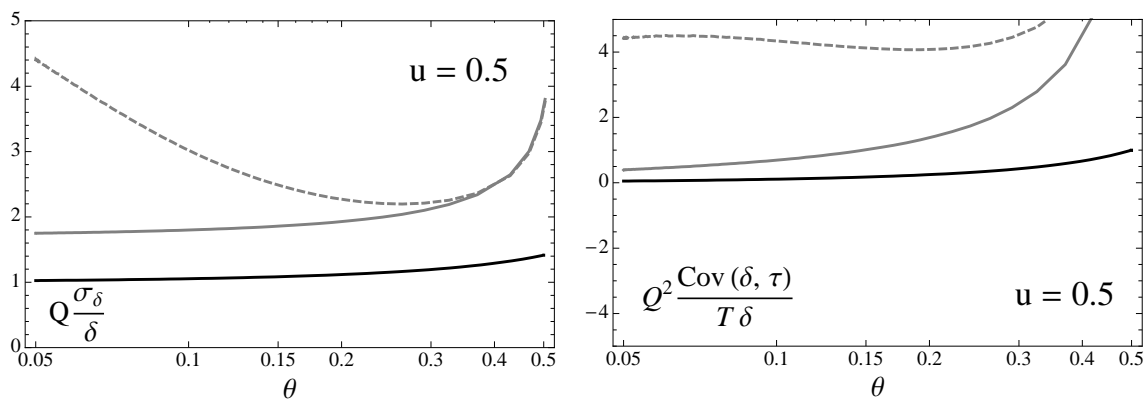}
   \caption{Comparison of select covariance matrix elements for the
     piecewise-linear model (black curve) and a linear limb-darkened
     light curve ($u=0.5$; gray curves), as a function of $\RHO
     \equiv \TE/\TCR$.   The $\DEPTH$ parameter has been
     redefined as the minimum of the limb-darkened light curve, as
     approximated by Eqn.~(\ref{eq:depth_ld}), in the solid gray curve.   The dashed gray curve uses the initial $\DEPTH$ association, as defined in Eqn.~(\ref{eq:depth}).  Linestyles follow the conventions of Fig.~(\ref{fig:cov_numerical}).}
   \label{fig:cov_ld_05_fix} 
\end{figure}

It is possible to improve the agreement with the analytic formulas by
associating $\DEPTH$ with the minimum of the transit light curve,
rather than the square of the radius ratio.  Specifically, one
replaces the definition of Eqn.~(\ref{eq:mappings_p2}) with the new
definition
\begin{eqnarray}
  \DEPTH & = & \FO\R^2~\frac{9 - 8 \left(\sqrt{1-b^2}-1\right) u}{9-8 u }.
\label{eq:depth_ld}
\end{eqnarray}
For the previously-derived formulas to be valid, we must adopt a value
for $u$ based on other information about the parent star (its spectral
energy distribution and spectral lines, luminosity, etc.)~rather than
determining $u$ from the photometric data.
Fig.~(\ref{fig:corr_ld_05_fix}) shows the correlations resulting from
this new association, for the case $u = 0.5$.  Fig.~(\ref{fig:cov_ld_05_fix}) shows the improvement with this new association for the variance in $\DEPTH$ and the covariance between $\DEPTH$ and $\TE$, for the case $u = 0.5$. While this new
association improves on the agreement with the analytic covariances (particularly at low normalized impact parameter), a
disadvantage is that we no longer have a closed-form mapping from
$\{\DEPTH, \TCR, \TE\}$ back to the more physical parameters $\{r, b,
\TAU\}$.

 It should be noted that there is evidence that linear limb darkening may not adequately fit high-quality transit light curves relative to higher order models (\citet{brown01}, \citet{south08}).  A more complete analysis with arbitrary source surface brightness would minimally include quadratic limb darkening but is outside the scope of this discussion.  \citet{pal08} completes a complementary analysis to this one of uncertainties in the quadratic limb darkening parameters themselves.
 
\clearpage

\clearpage

\section{Errors in derived quantities of interest in the absence of limb darkening\label{sec:prop}}

The parameters $\{\TCENTER, \TE, \TCR, \DEPTH, \FO\}$ are preferred
mainly because they lead to simple analytic
formulas for their uncertainties and covariances. The values of these
parameters are also occasionally of direct interest. In particular,
when planning observations, it is useful to know the transit duration,
depth, and the predicted midtransit time. Of more direct scientific
interest are the values of the ``physical'' parameters, such as the
planetary and stellar radii, the orbital inclination, and the mean
density of the star. Those latter parameters also offer clearer {\it a
  priori}\, expectations, such as a uniform distribution in $\cos i$.

For affine parameter transformations ${\bf p} \mapsto {\bf p'}$ , we
may transform the covariance matrix ${\bf C}$ via the Jacobian ${\bf
J} = \frac{\partial {\bf p'}}{\partial {\bf p}}$ as
\begin{eqnarray}
	{\bf C}' = {\bf J}^{{\rm T}} {\bf C} {\bf J}~.
\label{eq:trans_cov}
\end{eqnarray}

Using Eqns.~(\ref{eq:mappings_p2}--\ref{eq:mappings_t2}), we may
calculate the Jacobian
\begin{eqnarray}
	\frac{\partial \{\TCENTER, b^2, \TAU^2, \R, \FO\}}{\partial \{ \TCENTER, \TE, \TCR, \DEPTH, \FO\}} & = &
 \left(
\begin{array}{lllll}
 1 & 0 & 0 & 0 & 0 \\
 0 & \frac{\R \TCR}{\TE ^2} & \frac{\TCR}{4 \R} & 0 & 0 \\
 0 & -\frac{\R}{\TE} & \frac{\TE }{4 \R} & 0 & 0 \\
 0 & -\frac{\TCR}{2 \FO \R \TE } & -\frac{\TCR \TE }{8 \FO \R^3} & \frac{1}{2 \FO \R} & 0 \\
 0 & \frac{p \TCR}{2 \FO \TE } & \frac{\TCR \TE }{8\FO \R} & -\frac{\R}{2 \FO} & 1
\end{array}
\right)
\label{eq:jacob_native}
\end{eqnarray}
between the parameters of the piecewise-linear model and the more
physical parameter set when limb darkening is negligible. Using this Jacobian, the transformed
covariance matrix is
\begin{eqnarray}
	{\rm Cov}'( \{b^2, \TAU^2, \R, \FO\}~, \{b^2, \TAU^2, \R, \FO\}) = \;\;\;\;\;\;\;\;\;\;\;\;\;\;\;\;\;\;\;\;\;\;\;\;\;\;\;\;\;\;\;\;\;\;\;\;\;\;\;\;\;\;\;\;\;\;\;\;\;\;\;\;\;\;\;\;\;\;\;\;\;\;\;\;\nonumber\\
 \frac{1}{Q^2} \left(
\begin{array}{llll}
	   \frac{24- \RHO  (4 (\RHO -3) \RHO +23)}{4 (1-\RHO ) \RHO ^3}\R^2 & \frac{24-\RHO  (23-4 (\RHO -2) \RHO
   )}{16 (1-\RHO ) \RHO }  \TCR^2 & \frac{2 \RHO  +1}{4 \RHO(1 - \RHO )} \R^2& 0\\
  \frac{24-\RHO  (23-4 (\RHO -2) \RHO ) }{16 (1-\RHO ) \RHO } \TCR^2& \frac{24- \RHO  (4
   (\RHO -1) \RHO +23)}{64 \R^2 (1-\RHO )}\RHO  \TCR^4 & \frac{  1-2 \RHO  }{16
   (1-\RHO )} \RHO\TCR^2& 0\\
  \frac{2 \RHO  +1}{4\RHO( 1 - \RHO )} \R^2& \frac{  1-2 \RHO  }{16 (1-\RHO )}\RHO \TCR^2&
   \frac{1}{4(1- \RHO) }\R^2 & 0\\
  0 & 0 & 0 & 0
\end{array}
\right) +\nonumber\\  \
\frac{\ETA}{Q^2}\left(
\begin{array}{llll}
    \frac{\left(1-2 \RHO  \right)^2}{4 \RHO ^2}\R^2 & \frac{1}{16} \left(1-4 \RHO ^2\right)
   \TCR^2 & \frac{ \left(1-2 \RHO \right)}{4 \RHO } \R^2 & \frac{\R^3 \left(1-2 \RHO
   \right)}{2 \RHO }  \FO\\
  \frac{1}{16} \left(1-4 \RHO ^2\right) \TCR^2 & \frac{\RHO ^2 \left(1+2 \RHO
   \right)^2}{64 \R^2} \TCR^4 & \frac{1}{16}  \RHO  \left(1+2 \RHO \right)
   \TCR^2 & \frac{1}{8} \R \RHO  \left(1+2 \RHO \right) \FO \TCR^2 \\
  \frac{  \left(1-2 \RHO \right)}{4 \RHO } \R^2& \frac{1}{16}  \RHO 
   \left(1+2 \RHO \right) \TCR^2 & \frac{1}{4} \R^2  & \frac{1}{2} \R^3 
   \FO \\
  \frac{\R^3 \left(1-2 \RHO \right)}{2 \RHO }  \FO & \frac{1}{8} \R \RHO  \left(1+2 \RHO \right) \FO
   \TCR^2 & \frac{1}{2} \R^3 \FO & \R^4 \FO^2
\end{array}
\right) \label{eq:native_cov_eta0}
\end{eqnarray}
where we have ignored the unmodified covariance elements involving
$\TCENTER$, and have kept only the leading-order terms in $r$ in the
$\ETA$-dependent matrix.  

The standard errors for other functions of the parameters,
$f(\{\PR_i\})$, can be found via error propagation, just as in
Eqn.~(\ref{eq:trans_cov}),
\begin{eqnarray}
	{\rm Var}[f(\{\PR_i\})] & = & \sum_i \sum_j {\rm Cov}(\PR_i, \PR_j) \frac{\partial f}{\partial \PR_i} \frac{\partial f}{\partial \PR_j}.
\end{eqnarray}
The results for several interesting and useful functions, such as the
mean densities of the star and planet, are given in Table
(\ref{tab:prop}).  For brevity, the results are given in terms of the
matrix elements of Eqn.~(\ref{eq:native_cov_eta0}).  Simplified expressions for covariance matrix elements
in the limit of $\ETA \rightarrow 0$, $\RHO$ small (plentiful out-of-transit data) and negligible limb darkening are given in
Table (\ref{tab:small}).

\clearpage
\newcommand{\cov}[2]{{\rm Cov}(#1,#2)/#1 #2}
\newcommand{\var}[1]{{\rm Var}(#1)/#1^2}
\newcommand{\varb}[1]{{\rm \bf Var}(#1)/#1^2}
\newcommand{\covb}[2]{{\rm \bf Cov}(#1,#2)/#1 #2}
\begin{table}[htbp]
\centering
\begin{tabular}{||l | l|| c |}\hline
{\bf Quantity} & {\bf Variance (Standard Error Squared)} &{\bf Notes}\\ \hline \hline
$R_{p} = \R R_{\star} $ & $R_p^2 \left[ \var{\R}+(\log M_{\star}/M_{\odot})^2 {\rm \bf Var}(x)\right] $ & 1\\ \hline
$R_{\star}/a =  (\gamma_1/\gamma_2) 2 \pi \TAU /P $ & $\frac{1}{4}(R_{\star}/a)^2 {\rm Var}(\TAU^2)/\TAU^4$& \\ \hline
$R_{p}/a = (\gamma_1/\gamma_2) 2 \pi \TAU \R/P $ & $(R_{p}/a)^2\left[\frac{1}{4} {\rm Var}(\TAU^2)/\TAU^4+\var{r}\right]$ &\\ \hline
$|b| =(\gamma_2^2/\gamma_1) |a \cos i/ R_{\star}|$ & $\frac{1}{4}{\rm Var}(b^2)/b^2 $ & \\ \hline
$\begin{array}{l}|\cos i| \\= (\gamma_1^2/\gamma_2^3)2 \pi \TAU\, |b|\, /P\end{array}$ & $\frac{1}{4} \cos^2 i \left[ {\rm Var}(b^2)/b^4+{\rm Cov}(\TAU^2, b^2)/\TAU^2 b^2 +{\rm Var}(\TAU^2)/\TAU^4\right]$& \\
\hline  
$\begin{array}{l}\rho_{\star} \\= (\gamma_2/\gamma_1)^3(3/8 G \pi^2) P /\TAU^3  \end{array}$ & 
	$ \frac{9}{4} \rho_{\star}^2 {\rm Var}(\TAU^2)/\TAU^4 $ & \\ \hline
$\begin{array}{l}\rho_{p} \\ = \gamma_2 (K_{\star} \rho_\star/\R^3 \sin i)(P/ 2 \pi G M_\star)^{1/3}\end{array}$ & $\begin{array}{l} \rho_{p}^2 \left[\frac{9}{4}{\rm Var}(\TAU^2)/\TAU^4+9 \var{r}+\frac{9}{2}\cov{r}{\TAU^2}\right.\\
	\;\;\left.+\frac{1}{4}(\cos i /b)^4{\rm Var}(b^2)-\frac{3}{4}(\cos i /b)^2{\rm Cov}(b^2,\TAU^2)/\TAU^2\right.\\
	\;\;\left.-\frac{3}{2}(\cos i/b)^2{\rm Cov}(b^2,r)/r+\varb{K_{\star}}\right]
	\end{array}$ & 2
	\\ \hline
$\begin{array}{l}g_{\star} \\ = (\gamma_2/\gamma_1)^3 R_{\star} P/(2 \pi \TAU^3) \end{array}$ &
	$g_{\star}^2 \left[ \frac{9}{4} {\rm Var}(\TAU^2)/\TAU^4 +(\log M_{\star}/M_{\odot})^2 {\rm \bf Var}(x)\right]$&1\\ \hline
$\begin{array}{l}g_p \\= (\gamma_2^3/\gamma_1^2) K_{\star} P/ (2 \pi \R^2 \TAU^2 \sin i)\end{array}$ &
$\begin{array}{l}
	g_p^2 \left[ {\rm Var}(\TAU^2)/\TAU^4 +4 \var{\R}+2 \cov{\R}{\TAU^2}\right.\\
	\;\;\left.+\frac{1}{4}(\cos i /b)^4{\rm Var}(b^2)-\frac{1}{2}(\cos i /b)^2{\rm Cov}(b^2,\TAU^2)/\TAU^2\right.\\
	\;\;\left. -(\cos i/b)^2{\rm Cov}(b^2,r)/r+ \varb{K_{\star}}\right] 
	\end{array}$&  2 \\
\hline 
\end{tabular}
\caption{Table of transit quantities and associated variances, in terms
  of the matrix elements given in Eqn.~(\ref{eq:native_cov_eta0}).  We have assumed that both the orbital period, $P$, and stellar mass, $M_{\star}$, are known exactly. We have defined the noncircular-orbit parameters $\gamma_1 \equiv 1+e \sin \omega$ and $\gamma_2 \equiv \sqrt{1-e^2}$ where $e$ is the eccentricity and $\omega$ is the argument of pericenter (see \S~\ref{sec:model} for a discussion of eccentric orbits).  Notes: (1) A mass-radius relation $R_{\star} \propto (M_{\star}/M_{\odot})^{x}$ is assumed; (2) We have assumed $i \gtrsim 80^\circ$ in simplifying the inclination dependence in the variance.
  Quantities in bold are not determined by the transit model and must
  be provided from additional information.  $K_{\star}$ is the semi-amplitude of the source radial velocity.    Terms have been arranged in order of relative importance
  with the largest in absolute magnitude coming first.  Refer to Table~(\ref{tab:small}) for matrix elements of Eqn.~(\ref{eq:native_cov_eta0}) for the case in which the planet is small, the out-of-transit flux is known precisely and limb darkening is negligible.}
\label{tab:prop}
\end{table}
\begin{table}[htbp]
\centering
\begin{tabular}{@{}|c|c|@{}}\hline
$\begin{array}{@{}lll@{}}
	Q^2{\rm Var}(\R)/ \R^2  & \approx & 1/4  \\
	Q^2 {\rm Var}(b^2)/b^4 & \approx & 6 \R^2/\RHO^3 b^4  \\
	Q^2{\rm Var}(\TAU^2)/\TAU^4 & \approx & 3/2 \RHO 
\end{array}$ &
$\begin{array}{@{}lll@{}}
	Q^2 {\rm Cov}(b^2, \TAU^2)/b^2 \TAU^2 & \approx & 6 \R/\RHO^2 b^2 \\
	Q^2 {\rm Cov}(b^2, \R)/b^2 \R & \approx & \R/4 \RHO b^2  \\
	Q^2 {\rm Cov}(\TAU^2, \R)/\TAU^2 \R & \approx & 1/16 
\end{array}$ \\ \hline
\end{tabular}
\caption{Covariance matrix elements from Eqn. (\ref{eq:native_cov_eta0}) in the limit $\ETA \rightarrow 0$ and $\RHO$ small for use in Table (\ref{tab:prop}).  These approximations are valid in the case in which the planet is small, the out-of-transit flux is known precisely and limb darkening is negligible.}
\label{tab:small}
\end{table}
\clearpage

\section{Optimizing parameter sets for fitting data with small limb darkening\label{sec:smaller_corrs}}

\begin{figure}[htbp] 
   \centering
   \plotone{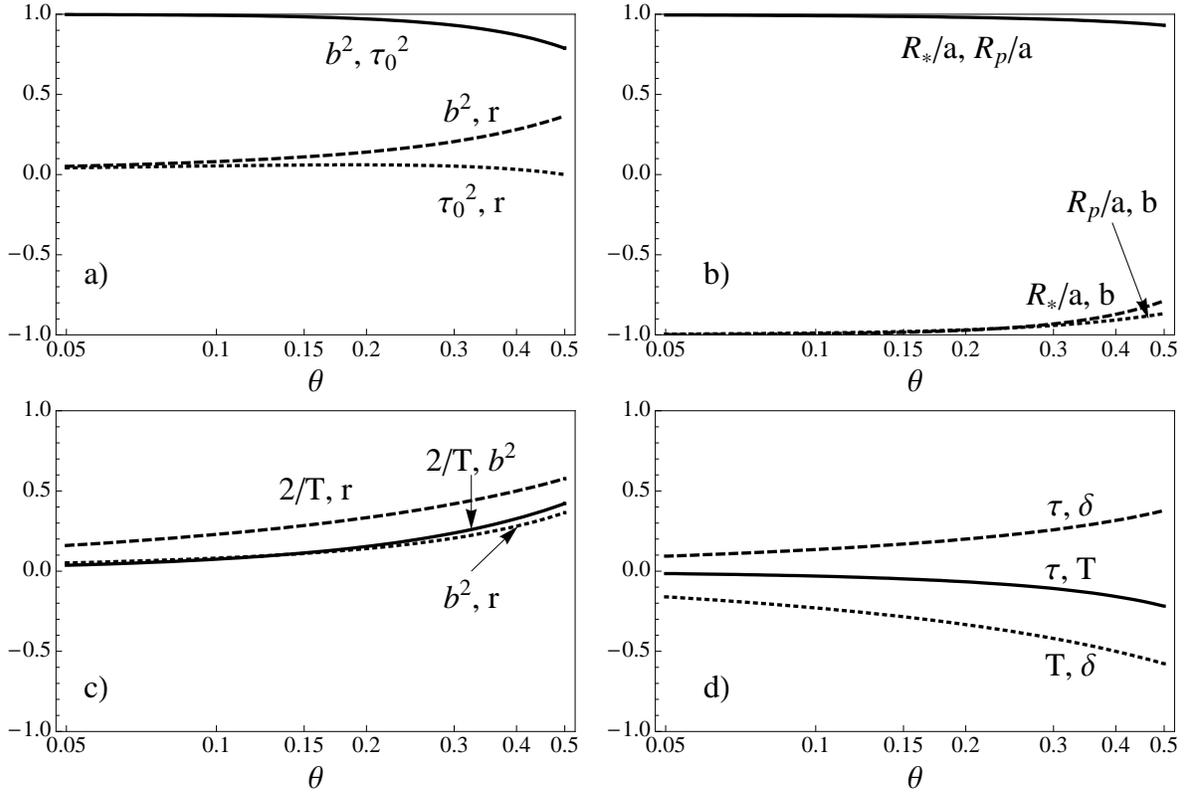}
   \caption{Comparison of correlations for various parameter sets that
   have been used in the literature. The correlations were derived from
   the piecewise-linear model (Eqn.~\ref{eq:sub_cov}) assuming $\ETA=0$.
   (a) Parameters $\{b^2, \TAU^2, \R\}$.
   (b) $\{R_{\star}/a = n \TAU, R_{p}/a = n \TAU \R, b^2\}$. 
   (c) $\{2/\TCR, b^2, \R\}$ (e.g., \citet{bakos07}).
   (d) $\{\TCR, \TE, \DEPTH\}$, the set introduced in this paper.}
   \label{fig:corr_other}
\end{figure}

The parameter set $\{\TCENTER, \TE, \TCR, \DEPTH, \FO\}$ has the
virtues of simplicity and weak correlation over most of the physical
parameter space.  However, when performing numerical analyses of
actual data, the virtue of simplicity may not be as important as the
virtue of low correlation, which usually leads to faster and more
robust convergence. To take one example, lower correlations among the
parameters result in reduced correlation lengths for Monte Carlo
Markov Chains, and faster convergence to the desired {\it a
  posteriori}\, probability distributions, and can obviate the need
for numerical Principal Component Analysis \citep{tegmark04}.  In
Fig.~(\ref{fig:corr_other}), we compare the degree of correlations for
various parameter sets that have been used in the literature on
transit photometry. Of note is the high degree of correlations among
the ``physical'' parameter set $\{R_\star/a, R_p/a, b\}$, which is a
poor choice from the point of view of computational speed.

\clearpage
\clearpage

Nevertheless, one advantage of casting the model in terms of physical
parameters is that the {\it a priori} expectations for those
parameters are more easily expressed, such as a uniform distribution
in $b$. The determinant of the Jacobian given by
Eqn.~(\ref{eq:trans_cov}), $|{\bf J}|$, is also useful in translating
{\it a priori}\, probability distributions from one parameter set to
the other [see \citet{burke07} or \citet{ford06} for an example of how this is done in
practice].  For the case of the parameter set $\{\TCENTER, \TE, \TCR,
\DEPTH, \FO\}$, we may use the Jacobian, Eqn.~(\ref{eq:jacob_native}),
to convert {\it a priori} probability distributions via

\begin{eqnarray}
	{\rm p}(\TCENTER, \TE, \TCR, \DEPTH, \FO) d\TCENTER\,d\TE\,d\TCR\,d\DEPTH\,d\FO& = &  {\rm p}(\TCENTER, b^2, \TAU^2, \R, \FO) \frac{1}{4\, \R\, \RHO\,\FO}d\TCENTER\,db^2\,d\TAU^2\,d\R\,d\FO \nonumber \\& = & {\rm p}(\TCENTER, b, \TAU, \R, \FO) \frac{1}{4\, \R\, \RHO\,\FO} \frac{1}{4\, b\, \TAU} d\TCENTER\,db\,d\TAU\,d\R\,d\FO\nonumber\\
	& = & {\rm p}(\TCENTER, b, \TAU, \R, \FO) \left(\frac{1-b^2}{16\, b\, \R^2 \,\TAU \,\FO}\right)d\TCENTER\,db\,d\TAU\,d\R\,d\FO.
\label{eq:remeasure}
\end{eqnarray}
where we have remeasured the phase space volume via the determinant,
\begin{eqnarray}
	\left|\left|\frac{\partial \{\TCENTER, b^2, \TAU^2, \R, \FO\}}{\partial \{ \TCENTER, \TE, \TCR, \DEPTH, \FO\}}\right|\right| & = &
	\frac{1}{4\, \R\, \RHO\, \FO}. \label{eq:det_jacob_native}
\label{eq:determinant}
\end{eqnarray}

\begin{figure}[htbp] 
   \centering
   \plotone{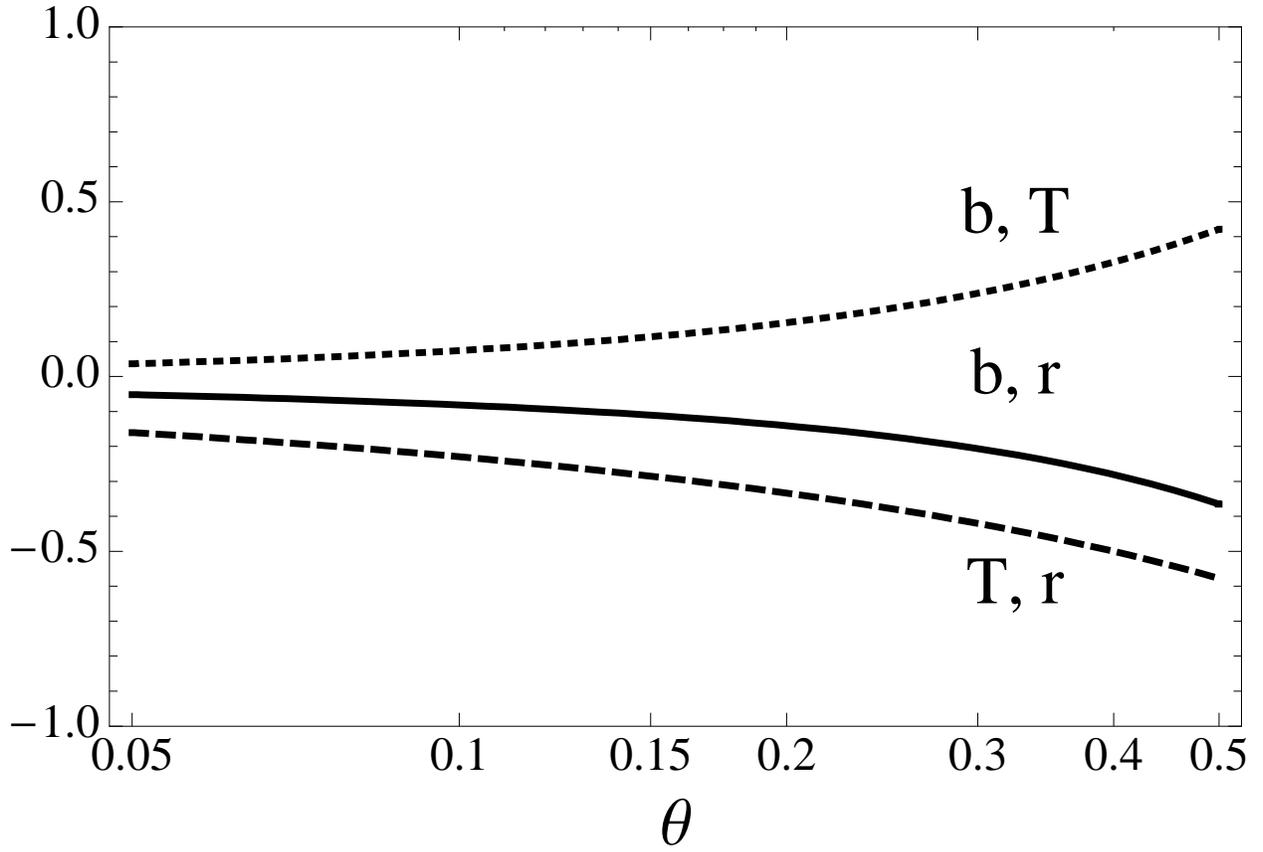}
   \caption{Correlations for the parameter set $\{b, \TCR, \R\}$.  The
     correlations were derived from the piecewise-linear model
     (Eqn.~\ref{eq:sub_cov}) assuming $\ETA = 0$.}
   \label{fig:uniform_priors}
\end{figure}

One may use this expression to enforce a uniform prior in $b$, for
example, by weighting the likelihood function as shown in
Eqn.~(\ref{eq:remeasure}). However, there is a practical difficulty
due to the singularity at $b = 0$. One way to understand the
singularity is to note that uniform distributions in $\TE$, $\TCR$
lead to a nearly uniform distribution in $\RHO = \TE/\TCR$, which
highly disfavors $b = 0$; in order to enforce a uniform distribution
in $b$, the prior must diverge at low $b$. Fig.~(\ref{fig:rho_in_b})
graphically captures the steep variation for small $b$ with $\RHO$.
Consider, instead, the parameter set $\{\TCENTER,b, \TCR, \R \equiv
\sqrt{\DEPTH/\FO}, \FO\}$ where, from Eqn.~(\ref{eq:mappings_b2}),
$b^2 = 1-\R \TCR/ \TE$.  We may calculate the determinant of the
Jacobian (not reproduced here)
\begin{eqnarray}
	\left| \left| \frac{\partial\{\TCENTER,b, \TCR, \R, \FO\}}{\partial \{\TCENTER, \TE, \TCR, \DEPTH, \FO\}} \right| \right| & = & \frac{(1-b^2)^2}{4 b~ \R^2 \FO \TCR}.
\end{eqnarray}
Combining this result with Eqn.~(\ref{eq:remeasure})
\begin{eqnarray}
{\rm p}(\TCENTER,b, \TCR, \R, \FO) d\TCENTER\,db\,d\TCR\,d\R\,d\FO& = &  {\rm p}(\TCENTER, b, \TAU, \R, \FO) \frac{1}{1-b^2}\frac{\TCR}{4\TAU} d\TCENTER\,db\,d\TAU\,d\R\,d\FO\nonumber\\ 
	& = &  {\rm p}(\TCENTER, b, \TAU, \R, \FO) \frac{1}{2 \sqrt{1-b^2}} d\TCENTER\,db\,d\TAU\,d\R\,d\FO. \label{eq:uprior}
\end{eqnarray}
The singularity at $b = 0$ has been removed with this parameter
choice.  There is a singularity at $b = 1$ instead, which is only
relevant for near-grazing transits, and is not as strong of a singularity
because of the square root.  We confirm that this parameter set also
enjoys weak correlations, as shown in Fig.~(\ref{fig:uniform_priors}),
and therefore this set is a reasonable choice for numerical
parameter-estimation algorithms. The merits of other parameter sets,
from the standpoint of correlation and {\it a priori} likelihoods, may
be weighed in a similar fashion, using the simple analytic covariance
matrix of Eqn.~(\ref{eq:cov}), and the appropriate transformation
Jacobian, in combination with Eqn.~(\ref{eq:trans_cov}).

\clearpage
\clearpage

\begin{figure}[htbp] 
   \centering
   \plotone{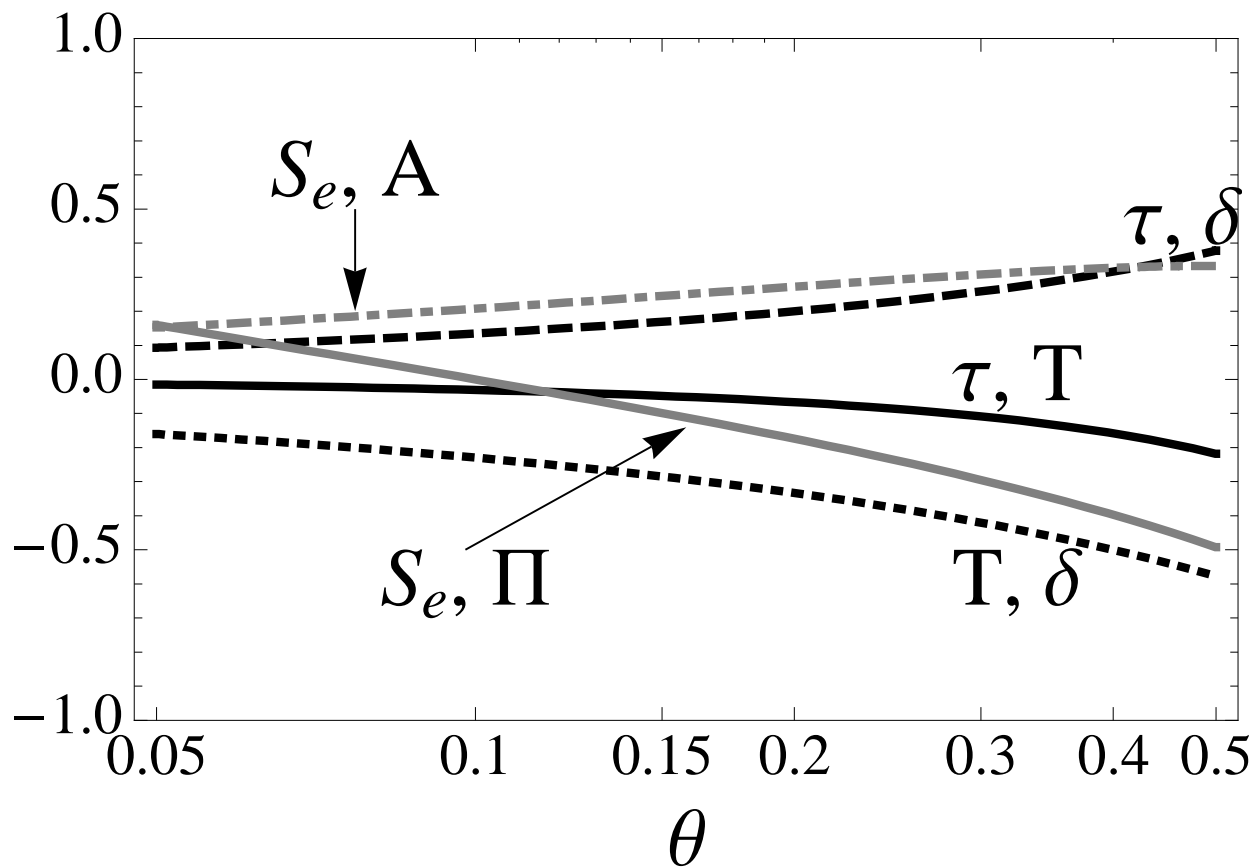}
   \caption{Comparison of the correlations among the parameters, for
     the set $ \{ \DEPTH, \TCR, \TE \} $ (black lines), the set $ \{ \SE, \TCR, \AREA = \TCR \DEPTH \}$ (dashed-dot gray line) and the set $ \{ \SE,
     \PI \equiv \TCR \DEPTH^{\THETAP}, \DEPTH \} $ (solid gray line) for the
     case $\THETAP=0.1$. For the latter set, the only nonzero
     correlation is between $\PI$ and $\SE$, which vanishes at $\RHO=0.1$. }
   \label{fig:param_comp}
\end{figure}

If the issues associated with the transformation of priors are ignored
(i.e.~if the data are of such quality that the results will depend
negligibly on the priors), we can give essentially uncorrelated
parameter sets. Consider, for example, the parameter set $\{ \TCENTER,
\SE \equiv \DEPTH/ \TE, \TCR, \AREA \equiv \DEPTH \TCR\}$.  The new
parameter $\SE$ is the magnitude of the slope of the light curve
during the ingress and egress phases, and the new parameter $\AREA$ is
the area of the trapezoid defined by the transit portion of the light
curve (i.e., the time integral of the flux decrement).  For simplicity
we assume $\ETA = 0$ and fix $\FO = 1$.  The transformed correlation
(Eqn. (\ref{eq:sub_corr})) is found via the transformation Jacobian,
Eqn. (\ref{eq:trans_cov}) as

\begin{eqnarray}
	{\rm Corr}(\{\TCENTER, \SE, \TCR, \AREA\}~ ,\{\TCENTER, \SE, \TCR, \AREA\}) &=&\left(
\begin{array}{llll}
 1 & 0 & 0 & 0\\
 0 & 1 & 0 & 0\\
 0 & 0 & 1 & \sqrt{\frac{\RHO (1-\RHO)}{(2-\RHO )  (\RHO +1)}}\\
 0 & 0 & \sqrt{\frac{\RHO (1-\RHO)}{(2-\RHO )  (\RHO +1)}} & 1
\end{array}
\right)
\label{eq:corr_newset}
\end{eqnarray}
The determinant of the transformation Jacobian (for use with Eqn.~(\ref{eq:remeasure})) is given as
\begin{eqnarray}
	\left| \left| \frac{\partial\{\TCENTER, \SE, \TCR, \AREA\}}{\partial \{\TCENTER, \TE, \TCR, \DEPTH\}} \right| \right| & = & \frac{(1-b^2)^2}{\TCR}.
\end{eqnarray}
With this new parameter set, the only nonzero correlation is between
$\TCR$ and $\AREA$, and this correlation is $\lesssim 0.3$ even for
grazing transits (see Fig.~\ref{fig:param_comp}). We have found that
these parameters provide a nearly optimal set for data fitting when
little is known at the outset about the impact parameter of the
transit.

It is possible to do even better when the impact parameter is known at
least roughly.  Consider the parameter set $\{ \TCENTER, \SE, \PI =
\TCR \DEPTH^{\THETAP}, \DEPTH\}$ where $\SE$ is the slope of ingress,
and $\THETAP$ is a constant (whose chosen value will be discussed
momentarily). The new parameter $\PI$ has no simple physical
interpretation. We again assume $\ETA = 0$ and $\FO=1$. The
correlation matrix in this case is
\begin{eqnarray}
	{\rm Corr}(\{\TCENTER, \SE, \PI, \DEPTH\}~ ,\{\TCENTER, \SE, \PI, \DEPTH\}) &=&\left(
\begin{array}{llll}
 1 & 0 & 0 & 0\\
 0 & 1 & 0 & 0\\
 0 & 0 & 1 &\frac{(\RHO-\THETAP)}{\sqrt{(\RHO-\THETAP)^2+2\RHO (1-\RHO)}}\\
 0 & 0 & \frac{(\RHO-\THETAP)}{\sqrt{(\RHO-\THETAP)^2+2\RHO (1-\RHO)}}  & 1
\end{array}
\right)
\label{eq:corr_2}
\end{eqnarray}
The determinant of the transformation Jacobian is given as
\begin{eqnarray}
	\left| \left| \frac{\partial\{\TCENTER, \SE, \PI, \DEPTH\}}{\partial \{\TCENTER, \TE, \TCR, \DEPTH\}} \right| \right| & = & \frac{(1-b^2)^2 \R^{2\THETAP}}{\TCR^2}.
\end{eqnarray}
With this choice, the only nonzero correlation is between $\PI$ and
$\DEPTH$. If the constant $\THETAP$ is chosen to be approximately
equal to $\RHO$, then this sole correlation may be nullified.  Thus,
if $\RHO$ is known even approximately at the outset of data
fitting---from visual inspection of a light curve, or from the
approximation $\RHO \approx \R$ valid for small planets on non-grazing
trajectories---a parameter set with essentially zero correlation is
immediately available.  As an example, Fig.~(\ref{fig:param_comp})
shows the correlation between $\PI$ and $\DEPTH$ as a function of
$\RHO$, for the choice $\THETAP = 0.1$, which has a null at $\RHO=0.1$
as expected.

The utility of this parameter set is not lost if $\THETAP$ cannot be confidently specified when used with Markov chain Monte Carlo parameter estimation codes.  At each chain step $i$, the next candidate state can be drawn from the candidate transition probability distribution function generated by the above parameter set with $\THETAP = \RHO_{i-1}$.  Thus, the Markov chain will explore the parameter space moving along principal axes at each chain step.  Additionally, allowing the candidate transition function to vary as the Markov chain explores parameter space may prove useful for low S/N data sets.

\clearpage
\clearpage

As a concrete example of the effectiveness of uncorrelated parameters,
we apply the MCMC algorithm to simulated data.  For a given choice of
the parameter set, we generate chains with a fixed jump-success
fraction, and calculate the resulting autocorrelations of the Markov
chain.  For a particular parameter $\PR$ (with value $\PR_i$ at chain
step $i$), the autocorrelation $a$ at a given chain step $j$ is defined
as
\begin{eqnarray}
	a_j & = & \frac{\langle \PR_i \PR_{i+j}\rangle - \langle \PR_i \rangle^2}{\langle \PR_i^2 \rangle-\langle \PR_i\rangle^2}
\label{eq:auto}
\end{eqnarray}
where the averages refer to the averages over the whole chain
\citep{tegmark04}.  The correlation length of the chain is the number
of steps $N$ that are required before the autocorrelation drops below
$0.5$.  The total chain length divided by the correlation length is
referred to as the effective length of a chain.  The effective chain
length is approximately the number of independent samples, which
quantifies the degree of convergence of the algorithm.  A lower
correlation length, for the same total chain length, gives a more
accurate final distribution. This autocorrelation analysis was
performed for both the ``physical'' parameter set $\{\TCENTER, b^2,
\TAU^2, r^2\}$ as well as the parameter sets $\{\TCENTER, \TE, \TCR,
\DEPTH\}$ and $\{\TCENTER, b, \TCR, \R\}$, with $\ETA=0$ in all cases
(i.e., plentiful out-of-transit data).  The MCMC was executed as
detailed in \S~\ref{sec:comp} with a fixed jump rate $\approx$50\% for
all parameter chains.  (In practice this was achieved by adjusting the
size of the Gaussian random perturbation that was added to each
parameter at each trial step.)  By choosing either the parameter set
$\{\TCENTER, \TE, \TCR, \DEPTH\}$ or $\{\TCENTER, b, \TCR, \R\}$, the
correlation lengths are reduced by a factor of approximately 150. By
using the minimally-correlated parameter set $\{\TCENTER, \SE, \TCR,
\AREA\}$, the correlation lengths are reduced by an additional factor
of $\sim$2.

To completely eliminate the correlations between parameters, one can
diagonalize the symmetric covariance matrix,
Eqn.~(\ref{eq:corr_newset}), and find the linear combinations of
parameters that eliminates correlations.  This was done by
\citet{burke07} for the particular case of the transiting planet
XO-2b. Analytic expressions for the eigenvectors are available because
there are only two entangled parameters. However, these eigenvectors
are linear combinations of local parameter values; they do not
constitute a global transformation rendering the covariance
diagonal. Thus, this procedure is useful for numerical analysis of a
particular system, although not for analytic insights.

\section{Summary}

We have presented formulas for uncertainties and covariances for a
collection of parameters describing the light curve of an exoplanet
transiting a star with uniform brightness. These covariances, given in
Eqns.~(\ref{eq:cov}, \ref{eq:native_cov_eta0}), are derived using a
Fisher information analysis of a linear representation of the transit
light curve. The key inputs are the uncertainty in each measurement of
the relative flux, and the sampling rate. We have verified the
accuracy of the variance and covariance estimates derived from the
piecewise-linear light curve with a numerical Fisher analysis of a
more realistic (nonlinear) light-curve model, and with a Markov Chain
Monte Carlo analysis of idealized data.

We focused on a particular parameterization of this piecewise-linear
light curve that we believe to be most useful. The parameters are the
midtransit time ($\TCENTER$), the out-of-transit flux ($\FO$), the flux decrement during
the full phase of the transit ($\DEPTH$), the duration of ingress or egress ($\TE$), and
the duration between the midpoint of ingress and the midpoint of
egress ($\TCR$).  This set is observationally intuitive and gives simple analytic
formulas for variances and covariances.  The exact parameter
definitions are provided in Eqns.~(\ref{eq:depth}, \ref{eq:tcr},
\ref{eq:te}) in terms of normalized impact parameter, stellar and
planetary radii, semi-major axis and orbital period.  Inverse mappings
to more physical parameters are provided in
Eqns.~(\ref{eq:mappings_p2}, \ref{eq:mappings_b2},
\ref{eq:mappings_t2}). The analytic covariance matrix is given in
Eqn.~(\ref{eq:cov}) and the analytic correlation matrix is given in
Eqn.~(\ref{eq:corr}).  Some quick-and-dirty (but still rather accurate) expressions
for the parameter uncertainties, for the case in which the planet is small,  the out-of-transit flux is known precisely and limb darkening is negligible, are given as
\begin{eqnarray}
	\sigma_{\TCENTER} & = & Q^{-1} \TCR \sqrt{\RHO/2},  \nonumber\\
	\sigma_{\TE}      & \approx & Q^{-1} \TCR \sqrt{6 \RHO},  \nonumber\\
	\sigma_{\TCR}     & \approx & Q^{-1} \TCR \sqrt{2 \RHO},  \nonumber \\
	\sigma_{\DEPTH}   & \approx & Q^{-1} \DEPTH \nonumber
\end{eqnarray}
where $\RHO \equiv \TE/\TCR$ is the ratio of the ingress or egress
duration to the total duration, and $Q \equiv \sqrt{\Gamma \TCR}
\frac{\DEPTH}{\sigma}$ is the total signal-to-noise ratio of the
transit in the small-planet limit (see
Eqn.~\ref{eq:dimensionless_vars}).

We investigated the applicability of these results to a limb darkened
brightness profile, in which the true light curve is not as
well-described by a piecewise-linear function.  We found that the
analytic formulas underestimate some of the variances and covariances
by a factor of a few, for a typical degree of limb darkening at
optical wavelengths. Significant improvements to covariance estimates
in the limb darkened case may be made by redefining the depth
parameter as a function of darkening coefficient and impact parameter
as in Eqn.~(\ref{eq:depth_ld}).  Unfortunately, no closed-form mapping
to more physical parameters exists with this choice, and therefore
most of the appeal of the analytic treatment is lost.

Quantities that are derived in part or in whole from the transit light
curve (such as stellar mean density or exoplanet surface gravity) are
provided in terms of the suggested parameter set.  In
Table~(\ref{tab:prop}), uncertainties propagated from the covariance
estimates for these quantities are provided with simple analytic
formulas.  In Table~\ref{tab:small}, covariance elements relevant to
the uncertainties in Table~\ref{tab:prop} are given for the case in
which the planet is small and the out-of-transit flux is known
precisely.  This allows the uncertainty in a given physical parameter
to be predicted in advance of any data, bypassing the need for
time-consuming simulations. For transit surveys, these formulas may
also be useful in giving closed-form expressions for the expected
distributions for some of the key properties of a sample of transiting
planets.

In \S~\ref{sec:smaller_corrs}, with the tools provided, we approach
the question of what parameter sets are best suited to numerical
parameter estimation codes.  This question depends both on the level
of parameter correlation and the behavior of any {\it a priori}
likelihood functions. We advocated a parameter set that has the virtue
of both weak correlation and essentially uniform {\it a priori}
expectations: specifically, the parameters are the midtransit time,
the out-of-transit flux, the ratio of planetary to stellar radii
($R_{p}/R_{\star}$), the normalized impact parameter, and the duration
between the midpoint of ingress and the midpoint of egress.
Fig. (\ref{fig:uniform_priors}) graphically describes the parameter
correlations while Eqn. (\ref{eq:uprior}) gives the {\it a priori}
probability distribution. Finally, two parameter choices are given
that are less intuitive than the suggested set but that provide
smaller correlations, depending on information that may be inferred or
guessed prior to analysis.  Correlations may be tuned to zero with the
second parameter choice for a non-grazing transit and an estimate of
$R_{p}/R_{\star}$.  The resulting correlation matrices for both
parameter choices are given in Eqns.~(\ref{eq:corr_newset},
\ref{eq:corr_2}).  Lower correlations relate directly to more
efficient data fitting, as demonstrated by reduced correlation lengths
with a Markov Chain Monte Carlo method.

\acknowledgements 
We thank Philip Nutzman for helpful comments on an
early version of this draft, and in particular for pointing out the
consequences of the singularity in Eqn.~(\ref{eq:determinant}). 
We also thank the referee for helpful comments, and for suggesting the Markov chain technique for use with the parameter choices in Eqn.~(\ref{eq:corr_2}). Sara
Seager and Paul Joss also provided helpful comments. We are grateful
for support from the William S.~Edgerly Innovation Fund and from NASA
grant HST-GO-11165 from the Space Telescope Science Institute, which
is operated by the Association of Universities for Research in
Astronomy, Incorporated, under NASA contract NAS5-26555.

\end{document}